\def\lae{\mathrel{<\kern-1.0em\lower0.9ex\hbox{$\sim$}}}
\def\gae{\mathrel{>\kern-1.0em\lower0.9ex\hbox{$\sim$}}}

\documentclass[12pt,preprint]{aastex}

\slugcomment{Accepted for publication in ApJS}
\lefthead{{JORD\'AN ET AL.}}
\righthead{ACSVCS II: DATA REDUCTION PROCEDURES}

\begin{document}

\title{The ACS Virgo Cluster Survey II. Data Reduction Procedures\altaffilmark{1}}

\author{Andr\'es Jord\'an\altaffilmark{2,3}, 
John P. Blakeslee\altaffilmark{4},
Eric W. Peng\altaffilmark{2},
Simona Mei\altaffilmark{4},
Patrick C\^ot\'e\altaffilmark{2}, 
Laura Ferrarese\altaffilmark{2}, 
John L. Tonry\altaffilmark{5},
David Merritt\altaffilmark{6}, 
Milo\v{s} Milosavljevi\'c\altaffilmark{7,8}, and
Michael J. West\altaffilmark{9}}

\begin{abstract}
The ACS Virgo Cluster Survey is a large program to carry out
multi-color imaging of 100 early-type members of the Virgo Cluster
using the Advanced Camera for Surveys (ACS) on the {\it
Hubble Space Telescope}.  Deep F475W and F850LP images
($\approx$ SDSS $g$ and $z$) are being used to  study the central
regions of the program galaxies, their globular cluster systems, and
the three-dimensional structure of Virgo itself. In this paper, we
describe in detail the data reduction procedures used for the survey,
including image registration, drizzling strategies, the computation
of weight images, object detection, the identification of globular
cluster candidates, and the measurement of their photometric and
structural parameters.
\end{abstract}


\altaffiltext{1}{Based on observations with the NASA/ESA 
{\it Hubble Space Telescope}
obtained at the Space Telescope Science Institute, which is operated 
by the Association
of Universities for Research in Astronomy, Inc., under 
NASA contract NAS 5-26555}
\altaffiltext{2}{Department of Physics and Astronomy, Rutgers University, 
Piscataway, NJ 08854; andresj@physics.rutgers.edu, ericpeng@physics.rutgers.edu, 
pcote@physics.rutgers.edu, lff@physics.rutgers.edu}
\altaffiltext{3}{Claudio Anguita Fellow}
\altaffiltext{4}{Department of Physics and Astronomy, 
Johns Hopkins University, Baltimore, MD 21218; jpb@pha.jhu.edu, smei@pha.jhu.edu}
\altaffiltext{5}{Institute of Astronomy, University of Hawaii, 
2680 Woodlawn Drive, Honolulu, HI 96822; jt@ifa.hawaii.edu}
\altaffiltext{6}{Department of Physics, Rochester Institute of Technology,
84 Lomb Memorial Drive, Rochester, NY 14623; merritt@mail.rit.edu}
\altaffiltext{7}{Theoretical Astrophysics, California Institute of Technology,
Pasadena, CA 91125; milos@tapir.caltech.edu.}
\altaffiltext{8}{Sherman M. Fairchild Fellow}
\altaffiltext{9}{Department of Physics \& Astronomy, University of Hawaii, 
Hilo, HI 96720; westm@hawaii.edu.}

\section{Introduction}
\label{sec:intro}

The Virgo Cluster is the largest concentration of early-type galaxies in
the Local Supercluster. As such, it has played a central role
in understanding how these galaxies form and evolve in dense environments,
providing invaluable information on the extragalactic distance scale, the
nature of galactic nuclei, globular cluster systems, and the shape and
universality of the galactic luminosity function. Indeed, 
the population of early-type Virgo galaxies has been the target
of several thousand observations with the {\it Hubble Space Telescope} (HST)
since its launch in 1990.

In the eleventh HST observing cycle, we initiated 
the ACS Virgo Cluster Survey (C\^ot\'e et~al. 2004; hereafter Paper I), a 
program to acquire F475W ($\approx$ SDSS $g$) and F850LP
($\approx$ SDSS $z$) images for 100 early-type members of
Virgo using the {\it Advanced Camera for Surveys} (ACS; Ford
et~al. 1998). Paper I of this series describes the survey itself, including
a brief overview of the scientific goals, the selection of the
program galaxies and their ensemble properties, the choice of filters, the
field placement and orientation, and the anticipated sensitivity limits.

It is the aim of the ACS Virgo Cluster Survey to use a single, homogeneous
dataset to measure surface brightness fluctuation (SBF) distances, study the
globular cluster systems, and examine the central structure
of 100 early-type members of the Virgo Cluster.
Both the scope of the survey --- consisting of 500 ACS
images spread over 100 separate fields --- and its multiple scientific goals 
require an automated and flexible data reduction procedure. In
this paper, we describe the pipeline used in the reduction and analysis
of the ACS imaging for this survey. As discussed in Paper I,
the ACS Virgo Cluster Survey also includes a substantial coordinated parallel
component, consisting of WFPC2 imaging for 100 ``intergalactic" fields in
Virgo; a complete discussion of the reduction and analysis of these parallel
data will be given in a separate paper. Similarly, future papers in this 
series will present additional details on the isophotal, globular cluster
and SBF analyses.

\section{Data Reduction Procedures}
\label{sec:red}

\subsection{Observations}

The program (GO-9401) consists of ACS imaging for 100 confirmed members
of the Virgo Cluster having morphological types E, S0, dE, dE,N or dS0. The
program galaxies span a factor of $\approx$ 450 in blue luminosity:
i.e., $9.31 \le B_T \le 15.97$. The 21 brightest program
galaxies constitute a complete sample of Virgo galaxies brighter than
$B_T = 12$, while the full sample includes nearly half of all early-type
members of Virgo with $B_T < 16$.

Observations for each program galaxy were carried out within a single
orbit with HST, using the ACS Wide Field Channel (WFC). 
This camera consists of two butted 2048$\times$4096 CCD detectors
(15$\mu$m pixels) having spectral response in the range 0.35--1.05$\mu$m.
With a scale of 0.049 arcsec per pixel, the camera has a
202$^{\prime\prime}\times202^{\prime\prime}$
field of view. The center of each galaxy was positioned near the
WFC aperture, at pixel position (2096,~200) on the WFC1 detector, and then offset 
perpendicular to the gap between the WFC1 and WFC2 detectors. For the 12 brightest
galaxies, this offset was 10$^{\prime\prime}$. An offset of 5$^{\prime\prime}$
was applied to the remaining galaxies. 

\begin{figure}
\plotone{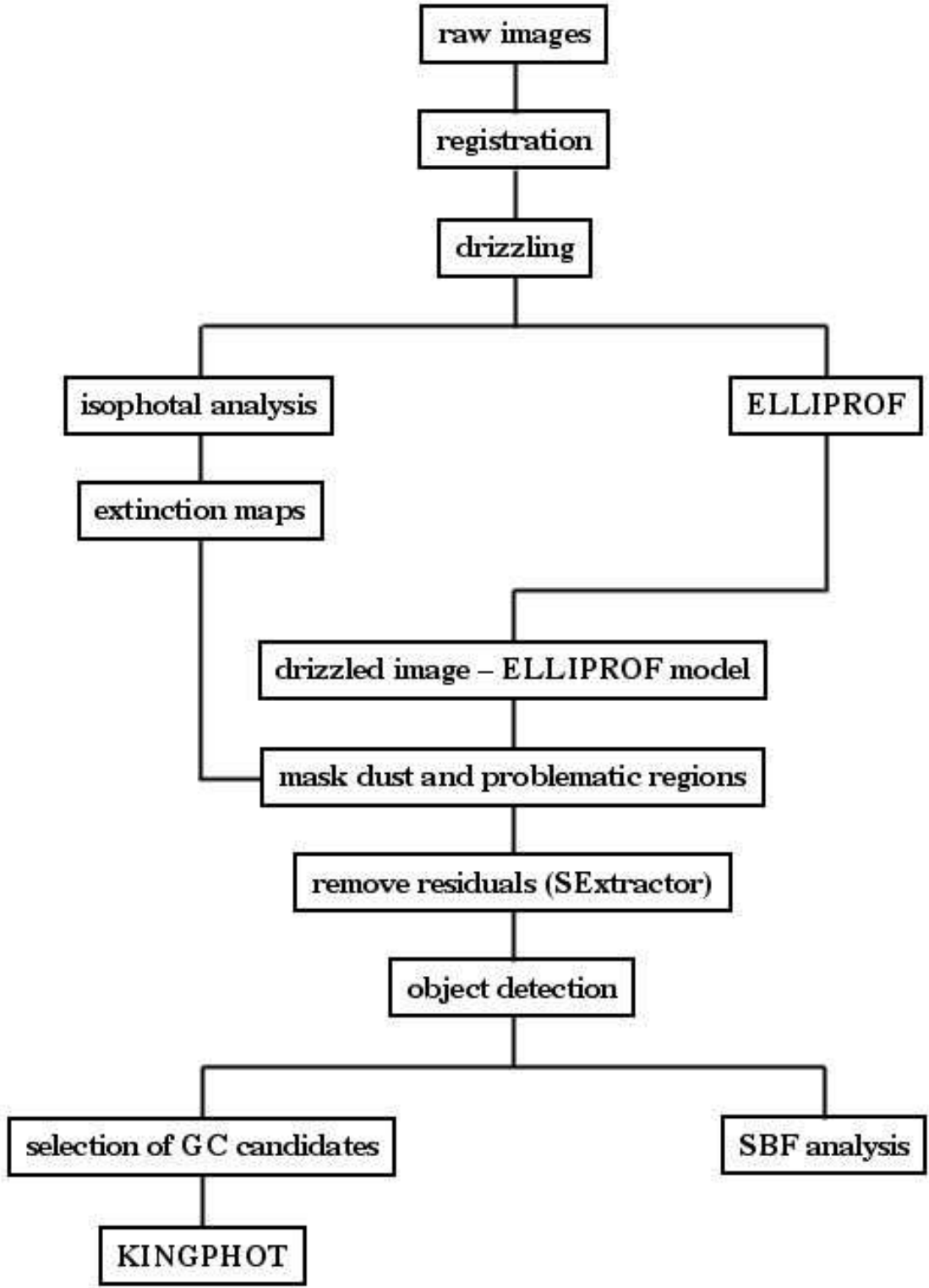}
\caption{Schematic representation of the data reduction pipeline for
the ACS Virgo Cluster Survey.
\label{fig01}}
\end{figure}

An identical observing sequence was adopted for each galaxy: $i.e.$, 
two 375 sec exposures in the F475W filter (750 sec in total), two 560 sec
exposures in the F850LP filter, and a single 90 sec exposure in F850LP
(1210 sec in total). For several of the program galaxies, the central
surface brightness in the redder bandpass can approach
${\mu}_z \simeq 12$~mag~arcsec$^{-2}$, so this 90 sec F850LP exposure
was required to repair nuclei saturated in the deep F850LP images.
The entire dataset for each galaxy therefore consists of an identical
set of five images, which were reduced and analyzed as described below.
These reduction procedures are summarized in schematic form by the
flowchart shown in Figure~\ref{fig01}.

Table~\ref{tab1} gives the observing log for all ACS observations related
to program GO-9401. From left to right, the columns of this table record
the identification number of each program galaxy, the Virgo Cluster Catalog
number from Binggeli, Sandage \& Tammann (1985), the universal date
of start of the observation, the dataset name, the universal time at
the start of the observation and the position angle, $\Theta$, of the
y axis of the WFC1 detector. The final two columns give the exposure
time and filter for each observation.

\subsection{Source Matching and Image Registration}

The imaging for each program galaxy was completed in
a single orbit --- at fixed telescope position and roll angle ---
so the images for each galaxy are expected to be closely 
registered. Nevertheless, small offsets might
be present, so shifts between the images were measured
using the method described in Blakeslee et al. (2003). A brief
overview of this method is given below.

To determine the shifts (${\Delta}x,~{\Delta}y$) and rotations between the
various images, SExtractor (Bertin \& Arnout 1996) was run with a signal-to-noise 
threshold of S/N = 10 on both science extensions for each image.\footnote{There
are two such extensions ($i.e.,$ CCD detectors) for the Wide Field
Channel on ACS.} The resultant catalogs were trimmed on the basis of object
size and shape parameters, thereby rejecting the vast majority of cosmic
rays, CCD artifacts, and diffuse extended objects.
If fewer than ten ``good'' sources remained, SExtractor
was rerun at a lower threshold. The ($x,y$) coordinates of each
source were then corrected using the distortion model read
from the IDCTAB FITS table specified in the image headers. For each
observation, sources in the two CCD chips were placed on a common 
rectified frame using the IDCTAB parameters V2REF and V3REF.

The MATCH program of Richmond et~al. (2000), which is based on the 
triangle-based search algorithm of Valdes et~al. (1995), was used
to derive shifts and rotations with respect to a reference image.
By default, this reference image was taken to be the one having the
greatest number of detected sources. The MATCH program was modified to
accept an input guess for the transformation (derived from the headers)
and report additional diagnostics to aid in evaluating the success of
the matching. The complete catalog of sources was
then used to fine tune this initial transformation.

This exercise revealed the measured rotations to be negligibly small
in all cases, so we simply evaluated the median (${\Delta}x,~{\Delta}$y)
shifts for the full list of matched sources.
The median shifts measured for the images were typically $\lae$ 0.2 pixel.

\subsection{Drizzling}

With the offsets determined in this way, the images were combined using the PYRAF
task \textit{multidrizzle} (Koekemoer et~al. 2002). Since several of the program galaxies
fill a significant fraction of the field of view ($i.e.$, the brightest galaxies
have effective radii $r_e \gtrsim 1^{\prime}$),
the task was run without sky subtraction.  Cosmic ray rejection was performed 
using this task; after some experimentation, the relevant parameters in the
task \textit{driz\_cr} were set to \texttt{scale = $^{\prime}$1.7 0.7$^{\prime}$} and
\texttt{snr = $^{\prime}$4.5 2.0$^{\prime}$}. The drizzled images have dimension
$4256\times4256$ pixels for all galaxies.

Given the differing requirements of the various scientific programs, both
\texttt{Lanczos3} and \texttt{Gaussian} kernels were used to distribute flux
onto the output image, thereby producing two independent image sets
for each galaxy. For the globular cluster and SBF analysis, images were
generated using a \texttt{Lanczos3} kernel, which reduces the correlation
between output pixels (and thus produces images with better noise
characteristics for SBF analysis). As a drawback, the negative lobes of
this kernel make the repair of bad pixels less effective. This is a potentially
significant problem for the measurement of surface brightness profiles in
the innermost regions of the program galaxies. Thus, the \texttt{Gaussian}
kernel was used to create the drizzled images used in the isophotal analysis
of the program galaxies.

In order to repair bad pixels, the \texttt{bits} parameter is set 
to indicate which pixels are to be considered good based on their
value in the data quality file and thus drizzled into the final image. 
For isophotal analysis, we set \texttt{bits=8322}, which allows for the
rejection of saturated pixels ($i.e.$, saturated pixels
are not drizzled and the final value in the drizzled image is obtained
by scaling based on the good pixels). Due to the difficulty in
repairing bad pixels with the \texttt{Lanczos3} kernel, we set
\texttt{bits=8576}, which allows the flux from saturated pixels to
be drizzled onto the final images.

\subsection{Weight Images}
\label{sec:wmap}

Weight images were constructed in order to perform object detection
and to aid in the determination of photometric and structural parameters
for the globular
cluster candidates (see \S\ref{sec:struc}). The presence of a strong
signal arising from the SBF of the program galaxies --- particularly
in the $z$ band --- requires that the ``noise" contributed by the
SBF must be included in the weight images in order to avoid the
detection of spurious sources corresponding to real fluctuations in
the underlying stellar populations. 

In constructing the weight images, we begin with the WHT images, 
$H_{ij}$, produced by {\it multidrizzle}, in which each pixel contains
the effective exposure time: $i.e.$, the total time during which
photons contributing to observed counts were collected.
The instrumental variance, $I_{ij}$, is then calculated as
\begin{equation}
I_{ij} = f_1 T + f_2 r^2   
\end{equation}
where $T$ is the exposure time and $r = 5e^{-}$ is the detector
readnoise for WFC/ACS. The factors $f_1$ and $f_2$ were
determined empirically by comparing the resulting variance images
calculated using the above expression with the noise measured
directly in exposures of varying duration drawn from the ACS Early Release
Observations. The values  determined in this way were $f_1=0.02222$ and 
$f_2=1.38$. With the instrumental noise properties in hand, weight
(inverse variance second$^{-2}$) 
images consisting of Poisson and readnoise were calculated as
\begin{equation}
W_{ij} = T H_{ij} \bigl ( T D_{ij} + I_{ij} \bigr ) ^{-1}
\end{equation}
where $D_{ij}$ denotes the data image in electrons second$^{-1}$. Weight
images in this form are then used in the measurement of photometric and
structural parameters for the globular cluster candidates, as described 
in \S\ref{sec:struc}.

As noted above, it is important to include the contribution of
SBF ``noise" when performing object detection. The ratio
of the variance from SBF, $\sigma^2_L$, to that of the 
Poisson noise from the galaxy, $\sigma^2_p$
is given by (Tonry \& Schneider 1988)
\begin{equation}
\frac{\sigma^2_L}{\sigma^2_p} = 10^{-0.4(\overline{m} - m_{\rm zpt*})}
\label{eq:sbfnoise}
\end{equation}
where 
$$\overline{m}  \equiv  -2.5\log(\sigma^2_L/ O_{ij} )+m_{zpt*}$$
is the fluctuation magnitude in the given band, 
$O_{ij} \equiv \sigma^2_p$ 
is the mean intensity of the galaxy, and 
$$m_{zpt*}=m_{zpt}+2.5\log (T).$$
Here $m_{zpt}$ is the
photometric zeropoint in the given band. From extensive
ground-based observations, the fluctuation magnitude of
bright Virgo ellipticals is known to be $\overline{I} \approx 29.6$
(Tonry et~al. 2001).
This value must be transformed to the bandpasses of ACS Virgo Cluster
Survey, bearing in mind that the correction in this case is
required for a typical program galaxy rather than a
luminous giant. To estimate this correction, we make use of
the scaling relation $\overline{I} \propto 4.5(V-I)$ from
Tonry et~al. (2001), and assume mean fluctuation colors of
$\langle \overline{z}_{850}-\overline{I} \rangle = -0.8$ and
$\langle \overline{g}_{475}-\overline{z}_{850} \rangle = 4$
based on the models
of Blakeslee, Vazdekis \& Ajhar (2001). From
Equation~\ref{eq:sbfnoise}, we then find
$\sigma^2_{L} \sim 1.2 O_{ij}$ and
$\sigma^2_{L} \sim 42 O_{ij}$ for the $g$ and $z$ bandpasses,
respectively. 
 
This estimate of the SBF contribution to the variance does not
take into account the effect of the PSF, which will
reduce its value per pixel. Convolving simulated noise images with
the ACS PSF, we found that the variance per pixel should be
reduced by factors of 12 and 14 for the $g$ and $z$ bandpasses,
respectively. 
With the galaxy model, $O_{ij}$, calculated
as described in \S\ref{sec:model} below,
the final weight image --- including Poisson noise, readnoise 
and the contribution from the SBF --- is then
\begin{equation}
W_{ij}^{\prime} = T H_{ij} \bigl ( T D_{ij} + I_{ij} + \kappa O_{ij} \bigr ) ^{-1}
\end{equation}
with $\kappa=0.1$ and 3 for the $g$ and $z$ bandpasses, respectively.
We emphasize again that this SBF ``noise" behaves in reality as a signal,
and by incorporating its contribution in the weight images, we are
effectively increasing the detection threshold in proportional
to the SBF signal. To include these weights in the object detection 
process, the SExtractor parameter \texttt{MAP\_RMS} was
set to $1/\sqrt{W^{\prime}_{ij}}$.

For some galaxies, internal dust obscuration, generally confined within
the core region, significantly affects the surface brightness profiles
and poses problems for both object detection and the estimation of the
local background. The method used to flag pixels affected by dust will
be described in detail in a future paper; very briefly, an optical 
depth map is created (in the $g$ band, for convenience) as 
$$A_{ij,g} = -2.5\times[\log(D_{ij,g}/D_{ij,z}) -\log C_{ij}]/(1-A_z/A_g).$$ 
This expression applies to the case in which the dust is located
entirely in the galaxy's
foreground; this assumption is almost certainly incorrect, and so
$A_{ij,g}$ represents a strict lower limit to the amount of absorption
present. In the dust regions, $C_{ij}$, the ratio of the unextincted
count rates in the $g$ and $z$ band, is estimated by linearly
interpolating (or extrapolating, if the dust affects the center) the
azimutally averaged values of $D_{ij,g}/D_{ij,z}$ observed in the
regions immediately surrounding the dust areas. A pixel is then flagged
as affected by dust if $A_{ij,g}$ is positive, larger than the local
standard deviation in the dust extinction map, and if more than one
contiguous pixel is affected. The last two conditions are necessary in
order not to flag noise spikes. In the case of galaxies that contain
large scale central dust disks, a more conservative approach was
followed prior to object detection and determination of the local
background, and the entire disk was masked by hand\footnote{
These galaxies are VCC1030, VCC1154, VCC1250 and VCC1535}.  
For both $W_{ij}$ and
$W_{ij}^{\prime}$, pixels that were
determined to have dust are given zero weight.
This prevents them from biasing the background determination (discussed
in Sec 2.5), creating false object detections, or biasing the fitting
of globular cluster surface brightness profiles. For VCC1316 (M87), 
pixels lying on its jet where also given zero weight.

\subsection{Galaxy Models and Object Detection}
\label{sec:model}

A two-dimensional model for each program galaxy, $O_{ij}$, is needed 
to compute the weight images used to perform object detection.
Once this model is in hand, it
is possible to subtract the galaxy and perform the detections
on a nearly flat background, greatly simplifying the estimation
of the background for the detected sources. This is particularly
important in the inner regions of the galaxies, where the
background can vary dramatically on small spatial scales.

\begin{figure}
\plotone{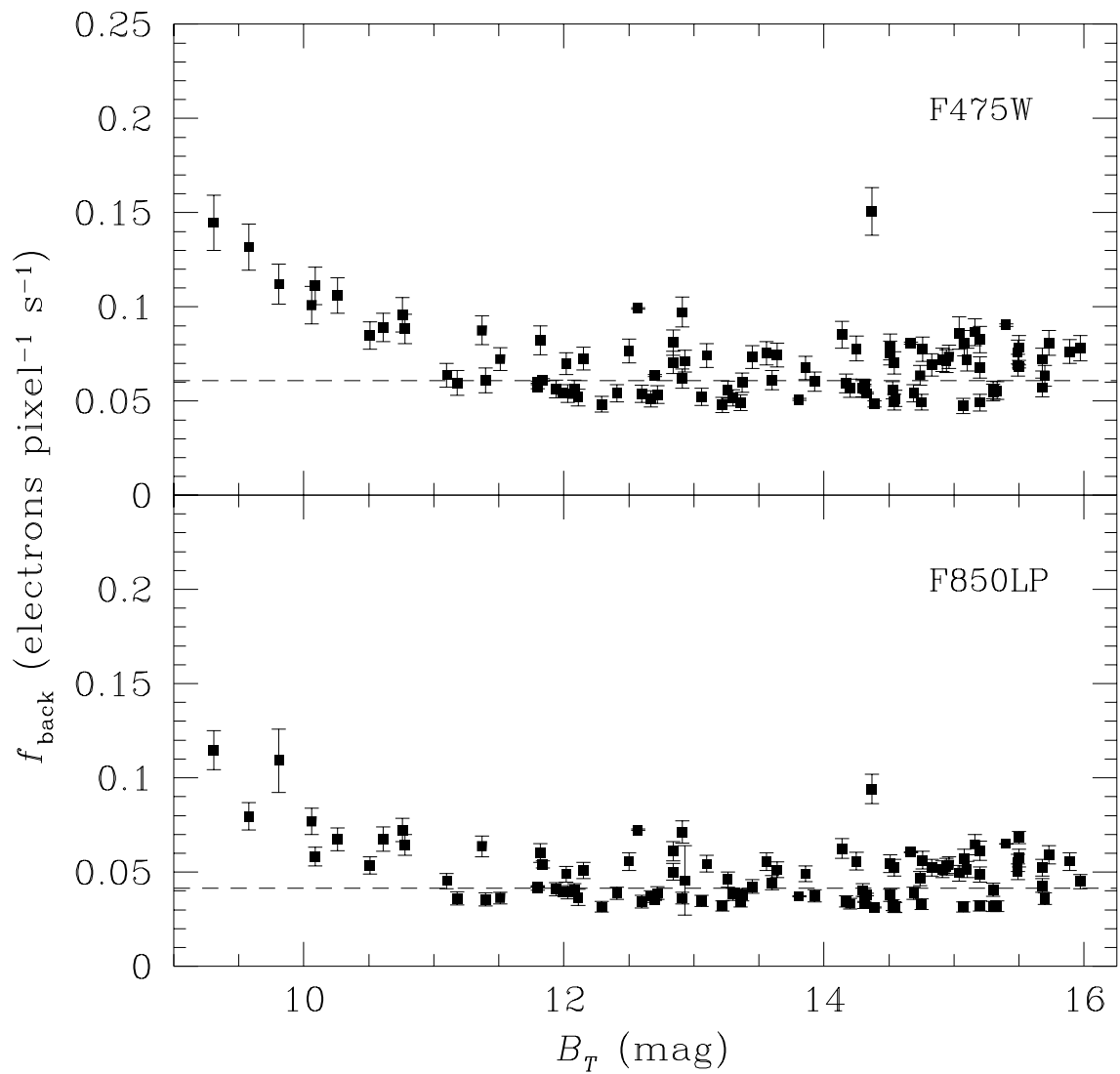}
\caption{Background count rates measured at $R \gtrsim 90^{\prime\prime}$
in the F475W and F850LP (upper and lower panels, respectively). The dashed lines 
in upper and lower panels show the count rates reported in the ACS Instrument
Handbook: $f_{\rm back}$ = 0.0609 and 0.0415 electrons pixel$^{-1}$ second$^{-1}$, respectively.
\label{fig02}}
\end{figure}

\begin{figure}
\plotone{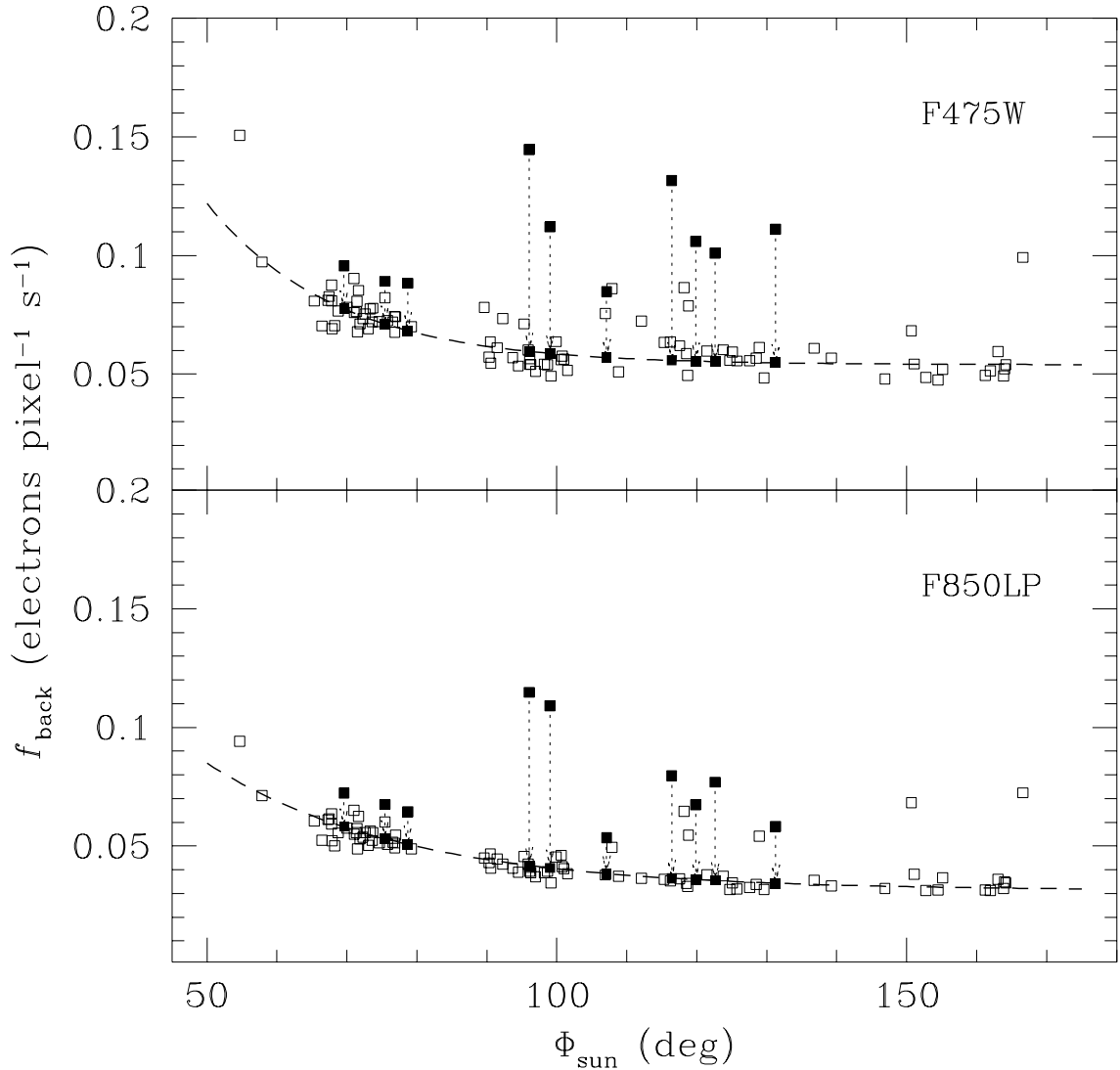}
\caption{Background count rates in F475W (upper panel) and F850LP (lower
panel) plotted as a function of $\Phi_{\rm sun}$. The dashed
curve in each panel shows a parametric representation of the form of
equation~5, determined directly from the 90 faintest galaxies in the
sample. The measured count rates for the 10 brightest galaxies are
shown by the upper filled squares; the adopted values are indicated by the
lower filled squares.
\label{fig03}}
\end{figure}

Galaxies were modeled using the ELLIPROF program
described in the SBF survey of Tonry et~al. (1997). This software
was used to estimate $O_{ij}$ and construct 
the weight images that include the contribution of the SBF
signal (\S\ref{sec:wmap}). A crucial preliminary step in 
construction of these models is the determination of the 
background ``sky" count rate for each galaxy. Figure~\ref{fig02} shows
the measured count rates, in both the F475W and F850LP filters,
for the full sample of galaxies.  These count rates, $f_{\rm back}$,
refer to the modal pixel values (in units of electrons pixel$^{-1}$
second$^{-1}$) measured at distances of $\approx$ 90--150\arcsec.
The upturn seen for the 10 brightest 
galaxies clearly reflects the contribution of the galaxies
themselves to the measured backgrounds ($i.e.,$ these galaxies have
a median effective radius of $\sim$ 50\arcsec). To estimate the true
background for these objects (all of which have similar locations
with respect to the ecliptic plane),
we use the fact that the background intensity scales with the SUNANGLE
parameter, $\Phi_{\rm sun}$, which gives the angle between the
Sun and the V1 axis of the telescope. Figure~\ref{fig03} shows
the measured background count rates as a function of $\Phi_{\rm sun}$
for the full sample of galaxies. The dashed curve in each
panel shows the function
$$f_{\rm back} = a_i e^{b_i(\Phi - c_i)} + d_i \eqno{(5)}$$
where the values of $a_i$, $b_i$, $c_i$ and $d_i$ for each filter
have been determined directly from the 90 faintest galaxies in the
sample. For the 10 brightest galaxies, which are shown as filled
squares in Figure~\ref{fig03}, we use equation~5 to estimate the
true background count rates.

While the ELLIPROF models generally
matched the galaxies quite well, there were in some cases
large-scale residuals which remained after subtracting the models. 
To remove these residuals, the 
following procedure was adopted. First, SExtractor 
was run on the image, $\Delta_{ij}$, constructed
by subtracting from the science image the galaxy model, 
$i.e.$, $\Delta_{ij}=D_{ij}-O_{ij}$. The SExtractor 
image is controlled by the parameters \texttt{BACK\_FILTERSIZE} and 
\texttt{BACK\_SIZE} which, after some experimentation,
were set to 1 and 40, respectively.\footnote{See Bertin \& 
Arnouts (1996) for a discussion of how the background is
estimated by SExtractor.}
The choice for the latter
parameter is particularly important since it must be set 
to a value small enough to remove much of the 
structure not accounted for by $O_{ij}$, yet
at the same time large enough to leave the power spectrum of the
point spread function (PSF) unaffected. The adopted value
of \texttt{BACK\_SIZE} = 40 accounts for structure on scales
of $\gtrsim 2\arcsec$ and yet is $\approx$ 20 times the FWHM of the PSF.

For seven of the galaxies, ELLIPROF was unable to produce acceptable models
because of the presence of strong disk components which were not well
approximated by elliptical isophotes modulated by low-order Fourier
terms.\footnote{These galaxies are VCC685, VCC1125, VCC1242, VCC1535,
VCC1692, VCC1857 and VCC2095.}
We experimented with parametric bulge+disk models, but these also resulted
in unacceptably large residuals.  Instead, we settled on the same basic
approach as used in the ground-based SBF survey for modeling early-type
disk galaxies: fitting 2-D bicubic splines after first taking the
logarithm of the galaxy image, except here used SExtractor to do the
fitting.  The logarithmic transformation reduces the steep surface
brightness gradients, allowing them to be fitted more accurately by the
spline interpolation.  We then took the inverse logarithm of the model,
subtracted it from the original image, and proceeded with fitting the
residuals in linear space as described above.

Object detection on the final subtracted image was performed using
a detection threshold of five connected pixels at a $1.5\sigma$
significance level.
The detections in both the F475W and F850LP images were then matched
using a matching radius of $0\farcs1$ which, after some experimentation,
was deemed to be optimal: $i.e.$, larger values produced an excess of
spurious matches, while some clear associations were missed for
smaller choices. A mask was constructed by assigning zero weight to
the elliptical regions of the object shape determined
by SExtractor and then scaling the $a,b$ parameters
by 1.2 times the Kron radius (in units of $a$). 
As the Kron radius will include $\sim 94\%$ of the object's flux, 
the factor of 1.2 ensures that most of the object's flux will be masked.
With this object mask, an improved background image, $\delta_{ij}$,
is then determined with SExtractor. By masking the objects when
estimating $\delta_{ij}$, we avoid biasing of the background by the
sources --- an important consideration given the rather small value
adopted for the \texttt{BACK\_SIZE} parameter.

The final F475W and F850LP images for object detection, photometry, and 
SBF analysis are then $F_{ij} = \Delta_{ij}-\delta_{ij}$. Object
detection was performed one last time by running SExtractor on
$F_{ij}$, with the detection performed
independently in both images, using a detection threshold
of five connected pixels at a $1.5\sigma$  significance level. Since
the background has already been subtracted from $F_{ij}$, SExtractor 
was run without further determination of the background ($i.e.$,
the background was fixed at a constant value of zero).

The pixel positions of the detections were converted to celestial
coordinates using the header information, and those sources detected
in a single filter only were discarded.  The detections in the F475W
and F850LP images were again matched using a matching radius of
$0\farcs1$. To investigate the accuracy of the absolute pointing, we 
compared the coordinates for 950 astrometric standards 
in our survey fields to those listed in the Guide Star Catalog 2.2
(McLean et~al. 1998). We find mean offsets of --0\farcs20 and
0\farcs18 in right ascension and declination, with rms scatter about
these value of 0\farcs57 and 0\farcs70, respectively. These results are
consistent with the absolute pointing accuracy of $\approx$ 0\farcs3 
reported in Jord\'an et~al. (2004; hereafter Paper III) based upon
a comparison of the celestial coordinates of the nucleus of VCC1316 
(M87 = NGC4486) with that measured using VLBI.
The internal accuracy of the celestial coordinates for a given galaxy
is $\approx$ 0\farcs01 (Meurer et~al. 2002).

\subsection{Globular Cluster Selection, Photometry and Structural Parameters}
\label{sec:struc}

Our final matched catalog of SExtractor sources consists of globular
clusters associated with each galaxy, as well as foreground stars and
compact background galaxies.  Additional selection criteria were 
applied to isolate candidate globular clusters for further analysis. 

\begin{figure}
\plotone{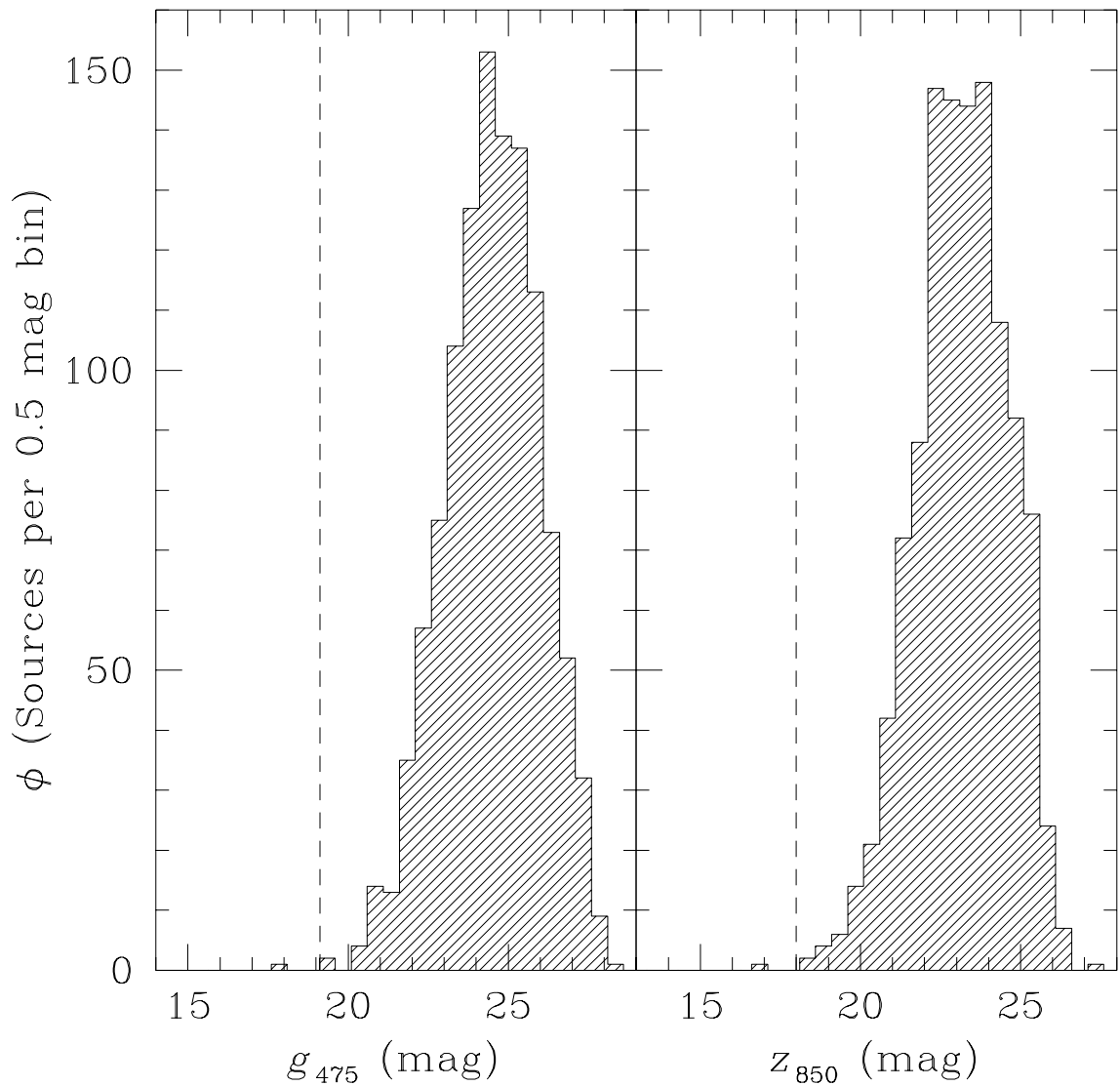}
\caption{{\it (Left Panel)} Luminosity function, $\phi$, in $g_{\rm 475}$ for
all 1144 sources detected in VCC1226 (M49 = NGC4472). The vertical dashed line
indicates the magnitude selection used to identify probable globular clusters.
{\it (Right Panel)} Same as the previous panel, except for $z_{\rm 850}$.
\label{fig04}}
\end{figure}

\begin{figure}
\plotone{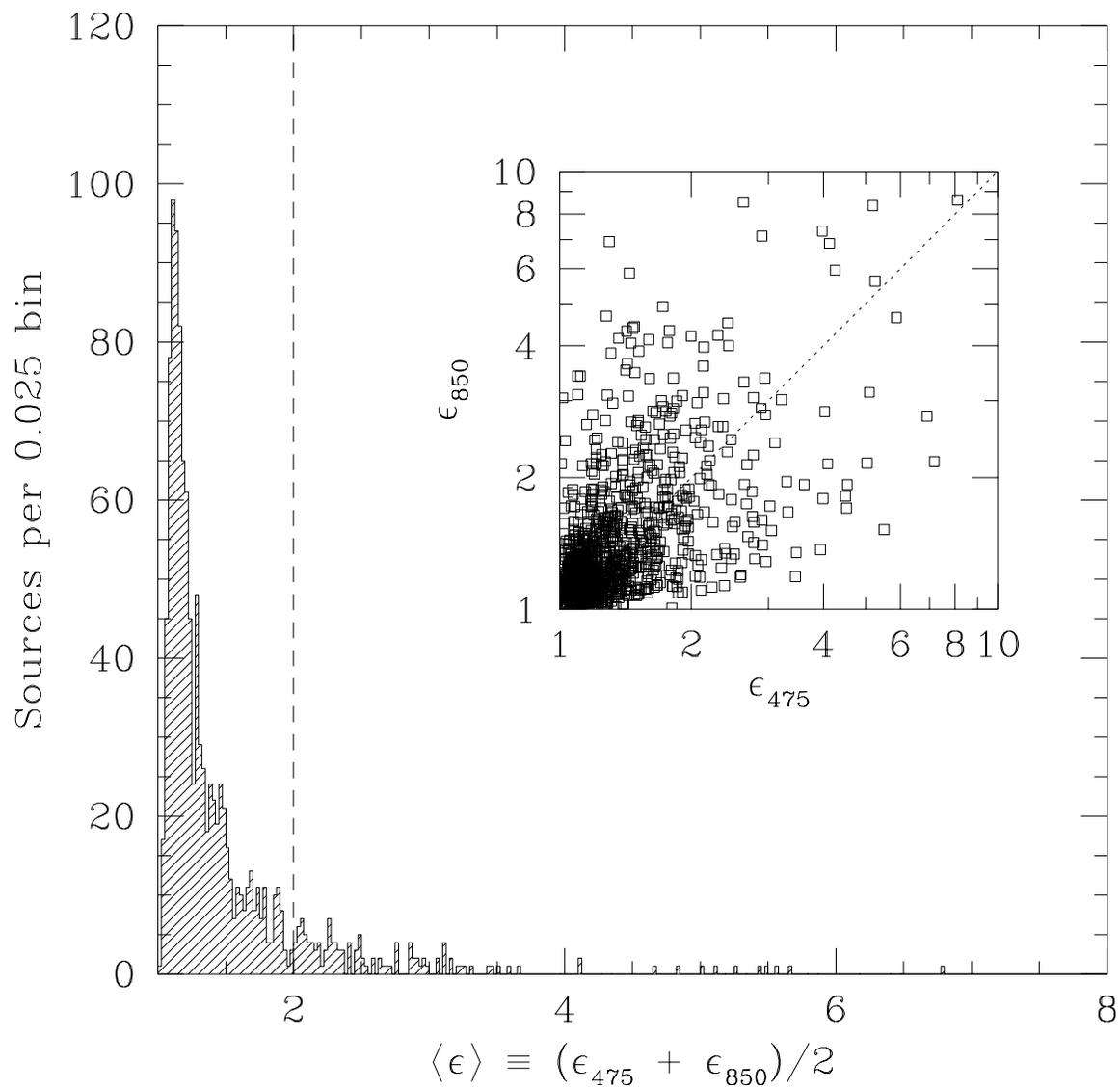}
\caption{Distribution of mean elongations, $\langle\epsilon\rangle$, for all
1144 sources detected in the field of VCC1226 (M49 = NGC4472). The vertical
dashed line indicates the selection on elongation, $\langle\epsilon\rangle < 2$,
used to identify probable globular clusters.  {\it (Inset)} Comparison
of the elongations measured in the F475W and F850LP images.
\label{fig05}}
\end{figure}

\begin{itemize}

\item[] {(1) {\underline{Selection on Magnitude}}}: It is
well established that the luminosity function of globular clusters
has a near Gaussian form ($e.g.$, Harris 2001). At the distance
of Virgo, the luminosity function peaks
at $V \approx 23.8$ (Whitmore et~al. 1995; Ferrarese,
C\^ot\'e \& Jord\'an 2003). To select a sample of
probable globular clusters, we discard all sources
in the SExtractor catalog with $g_{\rm 475} \le 19.1$ or
$z_{\rm 850} \le 18.0$. These limits are roughly five magnitudes 
brighter than the expected turnover of the globular cluster
luminosity function at the distance of Virgo.
For a luminosity function dispersion of
$\sigma = 1.40\pm0.05$ (Harris 2001), this selection 
will eliminate $\lesssim$ 0.02\% of the globular
clusters associated with the program galaxies, or only
$\sim$~2 clusters from the full survey. In any event, all
sources brighter than these cutoffs were inspected visually
to identify potentially interesting objects.
Figure~\ref{fig04} shows
the luminosity distribution of all sources in the SExtractor
object catalog for VCC1226 (M49 = NGC4472). The dashed vertical
lines in each panel indicate the magnitude cutoff used to select
probable globular clusters.

\item[] {(2) {\underline{Selection on Shape}}}: Globular 
clusters in the Local Group are nearly spherical systems, or
at most, only modestly flattened. We therefore discard those
sources in the SExtractor source catalog which have a mean
elongation, $\epsilon \equiv a/b$, measured in the 
F475W and F850LP filters to be $\langle \epsilon\rangle \ge 2$. This
generous limit easily includes even the most elongated clusters
in the Milky Way, M31 and the Magellanic Clouds (White \&
Shawl 1987; Lupton 1989; van den Bergh \& Morbey 1994).
The distribution of elongations for the full sample of sources
in VCC1226 is shown in Figure~\ref{fig05}. The inset to this
figure compares the elongations measured in the two bandpasses.

\end{itemize}

In addition to these cuts, any sources found within 10 pixels
($0\farcs5$) of the galaxy centers were omitted from the analysis
of the two-dimensional surface brightness profiles of the detected
sources (see below). In practice, this criterion eliminates only
the nuclei of the program galaxies. The properties of these nuclei,
and their relationship to the host galaxies, will be investigated 
separately in a future paper. 

At a distance of $\approx$~17~Mpc (Tonry et~al. 2001), 
the globular clusters associated with Virgo galaxies are
marginally resolved with WFC/ACS. For example, the mean
half-light radius of globular clusters in the Milky Way is
$\langle r_h \rangle \approx 3$~pc.
This translates into a half-light diameter of 0\farcs07,
or 1.45 pixels --- readily measurable given the 
$\approx 0\farcs1$ FWHM of the PSF in WFC mode. 

A Perl/PDL code (KINGPHOT) has been developed (Jord\'an \& C\^ot\'e
2004) to measure photometric and structural parameters
for those sources which satisfied the above criteria
by fitting the two-dimensional ACS surface brightness profiles with 
PSF-convolved isotropic, single-mass King (1966) models. This family
of models is well known to provide an excellent representation of 
the surface brightness profiles of Galactic globular clusters.
Empirical PSFs in both F475W and F850LP, varying quadratically
with CCD position, were derived using DAOPHOT II
(Stetson 1987; 1993) and archival observations of moderately
crowded fields in the outskirts of the Galactic globular
cluster $47$~Tucanae (NGC~104). The observations consisted of
a 30 sec F475W image from program GO-9656, and two 60 sec 
F850LP images from  program GO-9018. The archival images 
were drizzled in the same manner as the science images and a
total of $\approx$~200 stars were used in each case to
determine the PSF.

For each object classified as a globular cluster candidate,  KINGPHOT is
used to measure the total magnitude, King concentration index, $c$, and 
half-light radius, $r_h$ in both bandpasses. Note that we use
$r_h$ as scale factor in lieu of the more traditional core radius,
$r_c$, but the two parameters are related as described in
McLaughlin (2000). While the selection on magnitude and elongation
described above serves to lessen the contamination of the globular
cluster catalogs by stars and galaxies, some interlopers inevitably 
remain. In a future paper, we describe a method to further
reduce this contamination for each program galaxy by using our 
reduction pipeline to analyze archival F475W and F850LP blank
field images at high Galactic latitude.

\begin{figure}
\plotone{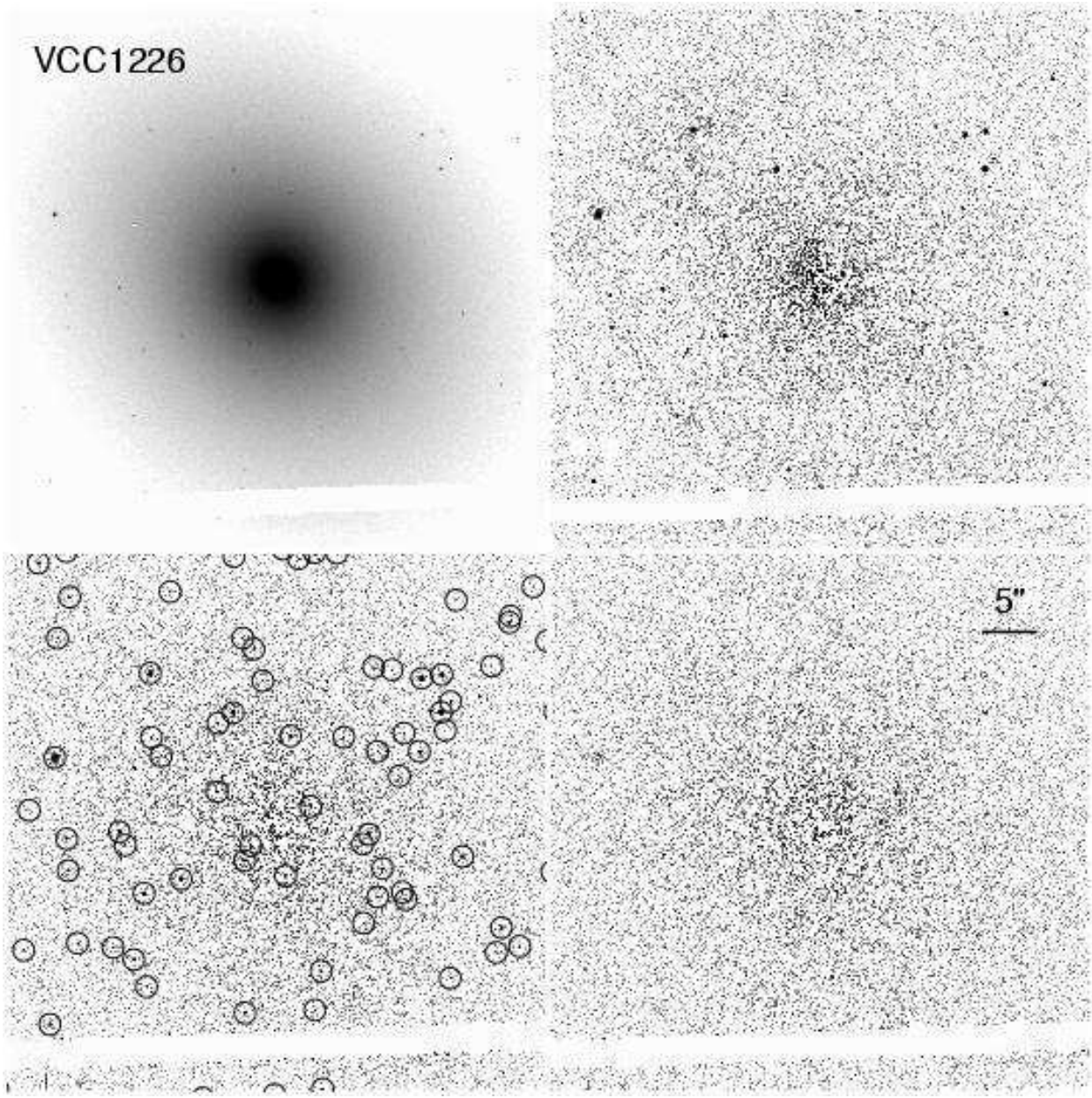}
\caption{An illustration of the various stages in the data reduction
pipeline for VCC1226 (M49 = NGC4472), the brightest galaxy in the ACS
Virgo Cluster Survey.  Each of the four panels shows a
$50^{\prime\prime}\times50^{\prime\prime}$ region centered on the nucleus.
{\it (Upper Left)} Drizzled and co-added F475W image.
{\it (Upper Right)} Image with ELLIPROF model for the galaxy subtracted. Note the
large-scale residuals.
{\it (Lower Left)} Image with SExtractor background removed, and sources classified
as globular cluster candidates identified.
{\it (Lower Right)} Image after the application of KINGPHOT, with the globular
cluster candidates subtracted from the previous image.
\label{fig06}}
\end{figure}

\begin{figure}
\plotone{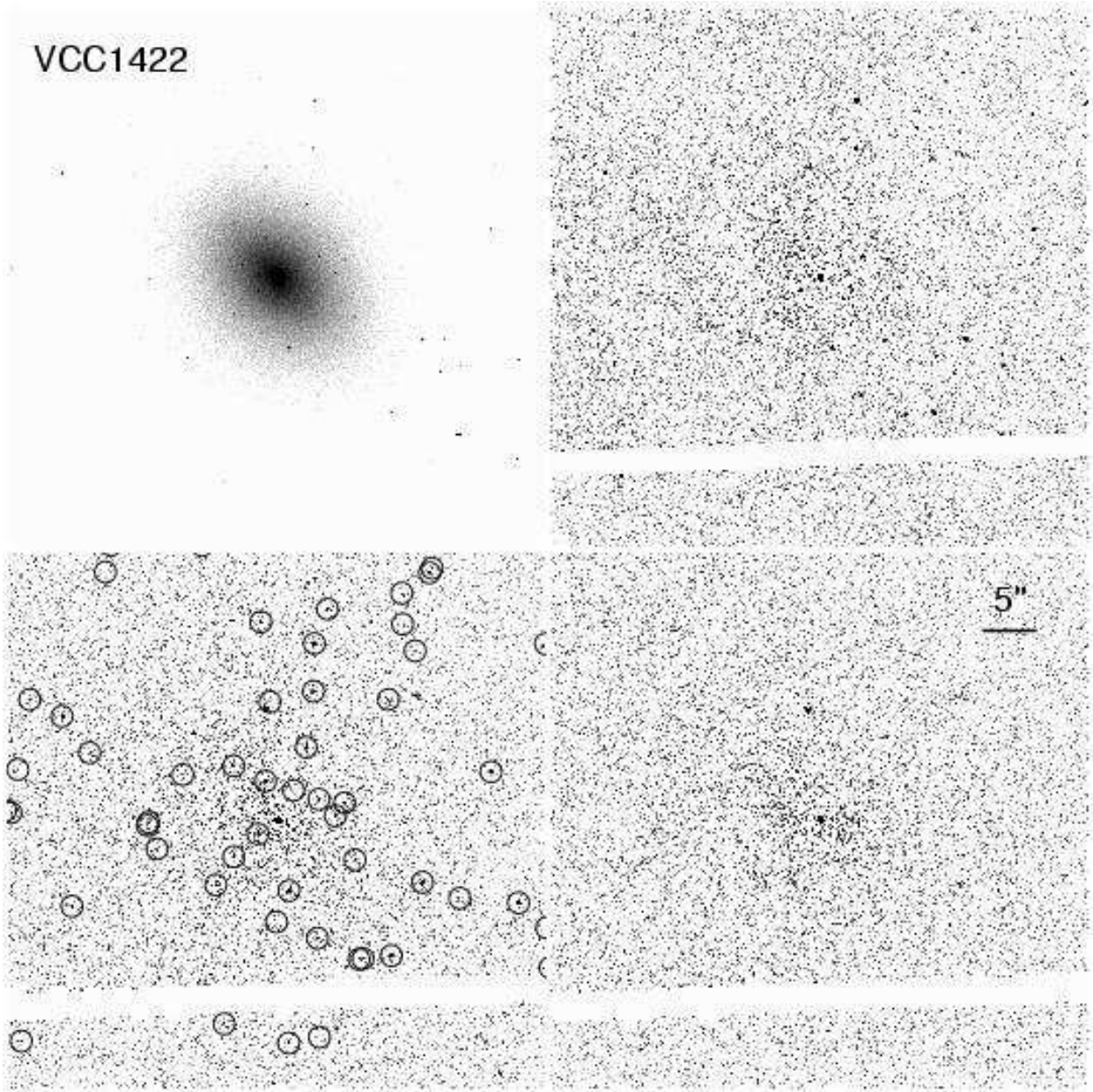}
\caption{Same as Figure~\ref{fig06}, except for VCC1422, the 50th brightest
galaxy in the survey.
\label{fig07}}
\end{figure}

\begin{figure}
\plotone{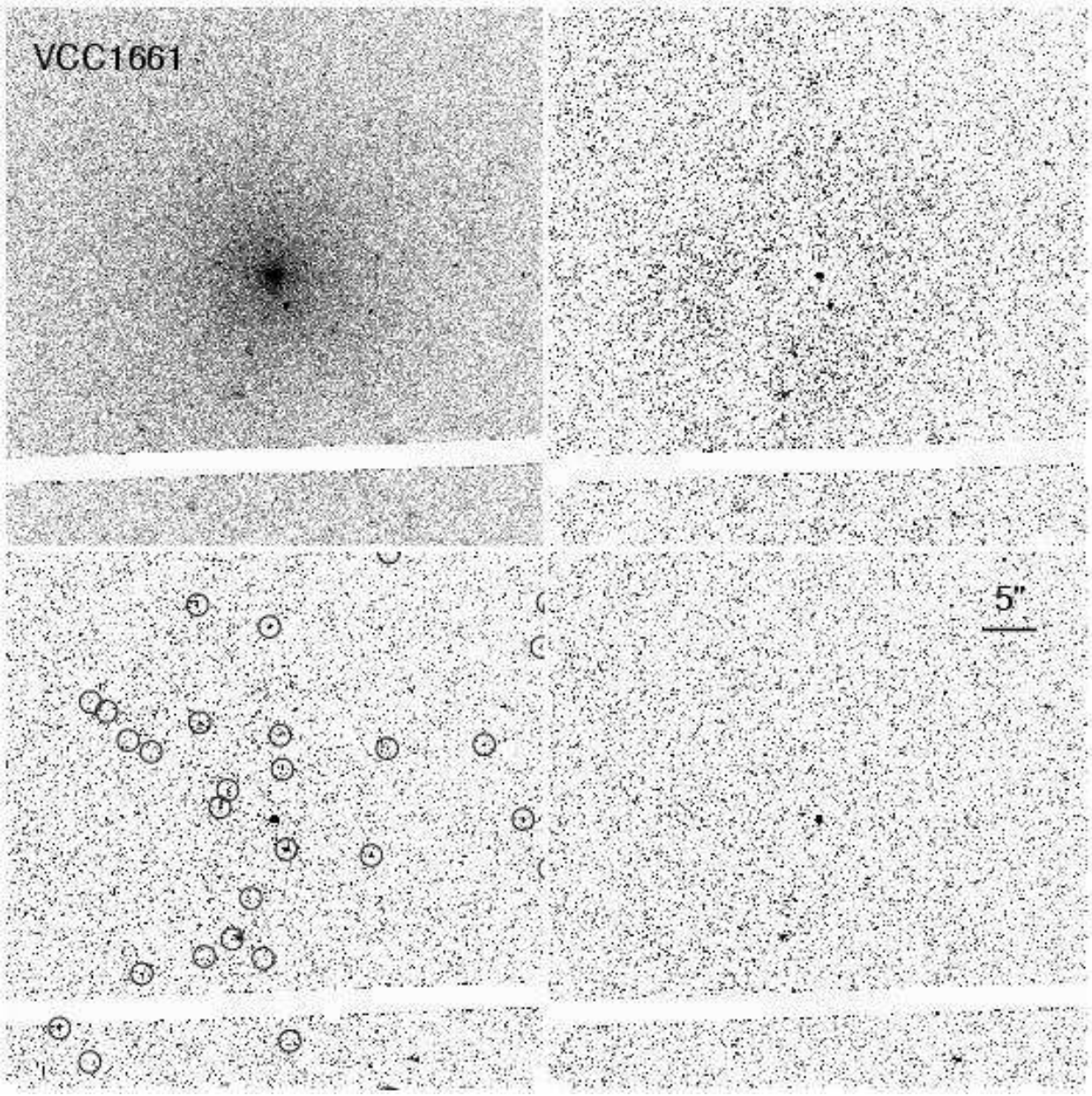}
\caption{Same as Figure~\ref{fig06}, except for VCC1661, the faintest of the
100 galaxies in the survey. 
\label{fig08}}
\end{figure}

Figures~\ref{fig06} -- \ref{fig08} illustrate the reduction procedures
described above for a representative sample of three program galaxies:
VCC1226, VCC1422 and VCC1661. These are the first, 50th and 100th
ranked galaxies in the survey, respectively. In each figure, we show
the registered, geometrically corrected F475W image in the upper left
panel. The upper right panel shows the residuals obtained 
after subtracting from this image the ELLIPROF model described
in \S\ref{sec:model}. The lower left panel shows the result of
subtracting the SExtractor model for this 
residual background; the circles in this panel indicate those
sources which satisfy the criteria used to select globular 
cluster candidates. Note that bright nuclei of VCC1422 and
and VCC1661 are excluded from the respective globular cluster
catalogs for these galaxies. The lower right panel shows the
image obtained by subtracting the best-fit, PSF-convolved King
model from each of the sources identified in the previous panel.

\subsection{Foreground Extinction and Photometric Calibration}

Since the Virgo Cluster spans $\approx$
10$^{\circ}$ on the sky, a reddening for each galaxy was 
computed using the DIRBE maps of Schlegel, Finkbeiner
\& Davis (1998). The mean reddening for the sample was
found to be $E(B-V)$ = 0.029 mag, with a
standard deviation of 0.008~mag. The corrections for
foreground extinction, assumed to be constant within each
ACS field, were taken to be
$$A_g = 3.634E(B-V)$$
$$A_z = 1.485E(B-V).$$
These extinction ratios 
correspond to the spectral energy distribution of a G2 star
(see Table~27 of Sirianni et~al. 2004), and are appropriate
for the globular clusters and elliptical galaxies targeted
in the survey.
Calibrated magnitudes on the AB system were
obtained using the relations
$$g_{\rm 475} = -2.5\log{(f_{\rm 475})} + 26.068$$
$$z_{\rm 850} = -2.5\log{(f_{\rm 850})} + 24.862$$
where $f_{\rm 475}$ and $f_{\rm 850}$ refer to the integrated
fluxes, in units of electrons/sec, in the F475W and F850LP filters.
The photometric zeropoints were taken from Sirianni et~al. (2004).

\section{Data Products}
\label{sec:conc}

The reduction pipeline described above was designed to meet the 
scientific objectives of the ACS Virgo Cluster Survey. Future papers
in this series will present scientific results from the
survey as well as a variety of data products. These 
products include a catalog of probable globular clusters
associated with each program galaxy and their basic properties
($e.g.$, celestial coordinates, magnitudes, colors
and structural parameters) and the results of an isophotal
analysis for each galaxy ($e.g$, radial profiles of
surface brightness, color, ellipticity, position angle,
as well as extinction maps and dust masses). These 
data products, along with raw and fully processed images used
in the analysis, will be made available through the project
website:
\url{http://www.physics.rutgers.edu/$\sim$pcote/acs}.

\acknowledgments

A.J. extends his thanks to the Oxford Astrophysics Department for their
hospitality.
Support for program GO-9401 was provided through a grant from the Space
Telescope Science Institute, which is operated by the Association of 
Universities for Research in Astronomy, Inc., under NASA contract NAS5-26555. 
A.J. acknowledges additional financial support provided by the National
Science Foundation through a grant from the Association of Universities
for Research in Astronomy, Inc., under NSF cooperative agreement
AST-9613615, and by Fundaci\'on Andes under project No.C-13442.
P.C. acknowledges additional support provided by NASA LTSA grant NAG5-11714.
D.M. is supported by NSF grant AST-020631, 
NASA grant NAG5-9046, and grant HST-AR-09519.01-A from STScI. 
M.M. acknowledges additional financial support provided by the Sherman
M. Fairchild foundation. 
M.J.W. acknowledges support through NSF grant AST-0205960.
This research has made use of the NASA/IPAC Extragalactic Database (NED)
which is operated by the Jet Propulsion Laboratory, California Institute
of Technology, under contract with the National Aeronautics and Space Administration.

\clearpage

\begin{deluxetable}{llllccll}
\tabletypesize{\scriptsize}
\tablecaption{Log of Observations for GO-9401.\label{tab1}}
\tablewidth{0pt}
\tablehead{
\colhead{ID} & 
\colhead{VCC} & 
\colhead{Date} & 
\colhead{Dataset} & 
\colhead{UT} &
\colhead{$\Theta$} &
\colhead{$T$} &
\colhead{Filter}  \\
\colhead{} &
\colhead{} &
\colhead{} &
\colhead{} &
\colhead{} &
\colhead{(deg)} &
\colhead{(sec)} &
\colhead{}  
}
\startdata
 1 &  1226  &  2003-06-19  &  j8fs01x5q  &  06:50:34  &  114.72  &   90  &  F850LP \\ 
   &        &    &  j8fs01x6q  &  06:54:22  &    &  560  &  F850LP \\ 
   &        &    &  j8fs01x8q  &  07:05:58  &    &  560  &  F850LP \\ 
   &        &    &  j8fs01xaq  &  07:18:11  &    &  375  &  F475W  \\ 
   &        &    &  j8fs01xcq  &  07:26:42  &    &  375  &  F475W  \\ 
 2 &  1316  &  2003-01-19  &  j8fs02bdq  &  20:47:35  &  269.99  &   90  &  F850LP \\ 
   &        &    &  j8fs02beq  &  20:51:23  &    &  560  &  F850LP \\ 
   &        &    &  j8fs02bgq  &  21:02:59  &    &  560  &  F850LP \\ 
   &        &    &  j8fs02biq  &  21:15:12  &    &  375  &  F475W  \\ 
   &        &    &  j8fs02bkq  &  21:23:43  &    &  375  &  F475W  \\ 
 3 &  1978  &  2003-06-17  &  j8fs03meq  &  18:02:11  &  118.60  &   90  &  F850LP \\ 
   &        &    &  j8fs03mfq  &  18:05:59  &    &  560  &  F850LP \\ 
   &        &    &  j8fs03mhq  &  18:17:35  &    &  560  &  F850LP \\ 
   &        &    &  j8fs03mjq  &  18:29:48  &    &  375  &  F475W  \\ 
   &        &    &  j8fs03mlq  &  18:38:19  &    &  375  &  F475W  \\ 
 4 &   881  &  2003-05-18  &  j8fs04y3q  &  08:09:39  &  140.20  &   90  &  F850LP \\ 
   &        &    &  j8fs04y4q  &  08:13:27  &    &  560  &  F850LP \\ 
   &        &    &  j8fs04y6q  &  08:25:03  &    &  560  &  F850LP \\ 
   &        &    &  j8fs04y8q  &  08:37:16  &    &  375  &  F475W  \\ 
   &        &    &  j8fs04yaq  &  08:45:47  &    &  375  &  F475W  \\ 
 5 &   798  &  2003-02-01  &  j8fs05dbq  &  21:00:42  &  276.94  &   90  &  F850LP \\ 
   &        &    &  j8fs05dcq  &  21:04:30  &    &  560  &  F850LP \\ 
   &        &    &  j8fs05deq  &  21:16:06  &    &  560  &  F850LP \\ 
   &        &    &  j8fs05dgq  &  21:28:19  &    &  375  &  F475W  \\ 
   &        &    &  j8fs05diq  &  21:36:50  &    &  375  &  F475W  \\ 
 6 &   763  &  2003-01-21  &  j8fs06o0q  &  20:49:42  &  285.65  &   90  &  F850LP \\ 
   &        &    &  j8fs06o1q  &  20:53:30  &    &  560  &  F850LP \\ 
   &        &    &  j8fs06o3q  &  21:05:06  &    &  560  &  F850LP \\ 
   &        &    &  j8fs06o5q  &  21:17:18  &    &  375  &  F475W  \\ 
   &        &    &  j8fs06o7q  &  21:25:50  &    &  375  &  F475W  \\ 
 7 &   731  &  2003-06-06  &  j8fs07f7q  &  13:08:41  &  130.00  &   90  &  F850LP \\ 
   &        &    &  j8fs07f8q  &  13:12:29  &    &  560  &  F850LP \\ 
   &        &    &  j8fs07faq  &  13:24:05  &    &  560  &  F850LP \\ 
   &        &    &  j8fs07fcq  &  13:36:18  &    &  375  &  F475W  \\ 
   &        &    &  j8fs07feq  &  13:44:49  &    &  375  &  F475W  \\ 
 8 &  1535  &  2003-07-12  &  j8fs08lnq  &  10:18:31  &  109.20  &   90  &  F850LP \\ 
   &        &    &  j8fs08loq  &  10:22:19  &    &  560  &  F850LP \\ 
   &        &    &  j8fs08lqq  &  10:33:55  &    &  560  &  F850LP \\ 
   &        &    &  j8fs08lsq  &  10:46:08  &    &  375  &  F475W  \\ 
   &        &    &  j8fs08luq  &  10:54:39  &    &  375  &  F475W  \\ 
 9 &  1903  &  2003-07-19  &  j8fs09lhq  &  07:09:57  &  106.61  &   90  &  F850LP \\ 
   &        &    &  j8fs09liq  &  07:13:45  &    &  560  &  F850LP \\ 
   &        &    &  j8fs09lkq  &  07:25:21  &    &  560  &  F850LP \\ 
   &        &    &  j8fs09lmq  &  07:37:34  &    &  375  &  F475W  \\ 
   &        &    &  j8fs09loq  &  07:46:05  &    &  375  &  F475W  \\ 
10 &  1632  &  2003-07-07  &  j8fs10eeq  &  10:17:18  &  110.94  &   90  &  F850LP \\ 
   &        &    &  j8fs10efq  &  10:21:06  &    &  560  &  F850LP \\ 
   &        &    &  j8fs10ehq  &  10:32:42  &    &  560  &  F850LP \\ 
   &        &    &  j8fs10ejq  &  10:44:55  &    &  375  &  F475W  \\ 
   &        &    &  j8fs10elq  &  10:53:26  &    &  375  &  F475W  \\ 
11 &  1231  &  2003-06-12  &  j8fs11a8q  &  14:47:10  &  116.84  &   90  &  F850LP \\ 
   &        &    &  j8fs11a9q  &  14:50:58  &    &  560  &  F850LP \\ 
   &        &    &  j8fs11abq  &  15:02:34  &    &  560  &  F850LP \\ 
   &        &    &  j8fs11adq  &  15:14:47  &    &  375  &  F475W  \\ 
   &        &    &  j8fs11afq  &  15:23:18  &    &  375  &  F475W  \\ 
12 &  2095  &  2003-03-21  &  j8fs12ipq  &  18:38:38  &  227.99  &   90  &  F850LP \\ 
   &        &    &  j8fs12iqq  &  18:42:26  &    &  560  &  F850LP \\ 
   &        &    &  j8fs12isq  &  18:54:02  &    &  560  &  F850LP \\ 
   &        &    &  j8fs12iuq  &  19:06:15  &    &  375  &  F475W  \\ 
   &        &    &  j8fs12iwq  &  19:14:46  &    &  375  &  F475W  \\ 
13 &  1154  &  2003-07-17  &  j8fs13xmq  &  03:56:34  &  108.61  &   90  &  F850LP \\ 
   &        &    &  j8fs13xnq  &  04:00:22  &    &  560  &  F850LP \\ 
   &        &    &  j8fs13xpq  &  04:11:58  &    &  560  &  F850LP \\ 
   &        &    &  j8fs13xrq  &  04:24:11  &    &  375  &  F475W  \\ 
   &        &    &  j8fs13xtq  &  04:32:42  &    &  375  &  F475W  \\ 
14 &  1062  &  2003-02-09  &  j8fs14knq  &  17:56:13  &  288.99  &   90  &  F850LP \\ 
   &        &    &  j8fs14koq  &  18:00:01  &    &  560  &  F850LP \\ 
   &        &    &  j8fs14kqq  &  18:11:37  &    &  560  &  F850LP \\ 
   &        &    &  j8fs14ksq  &  18:23:50  &    &  375  &  F475W  \\ 
   &        &    &  j8fs14kuq  &  18:32:21  &    &  375  &  F475W  \\ 
15 &  2092  &  2003-01-20  &  j8fs15g6q  &  22:27:38  &  287.21  &   90  &  F850LP \\ 
   &        &    &  j8fs15g7q  &  22:31:26  &    &  560  &  F850LP \\ 
   &        &    &  j8fs15g9q  &  22:43:02  &    &  560  &  F850LP \\ 
   &        &    &  j8fs15gbq  &  22:55:14  &    &  375  &  F475W  \\ 
   &        &    &  j8fs15gdq  &  23:03:46  &    &  375  &  F475W  \\ 
16 &   369  &  2003-06-14  &  j8fs16t2q  &  19:35:28  &  129.80  &   90  &  F850LP \\ 
   &        &    &  j8fs16t3q  &  19:39:16  &    &  560  &  F850LP \\ 
   &        &    &  j8fs16t5q  &  19:50:52  &    &  560  &  F850LP \\ 
   &        &    &  j8fs16t7q  &  20:03:05  &    &  375  &  F475W  \\ 
   &        &    &  j8fs16t9q  &  20:11:36  &    &  375  &  F475W  \\ 
17 &   759  &  2003-06-11  &  j8fs17pnq  &  08:22:38  &  112.81  &   90  &  F850LP \\ 
   &        &    &  j8fs17poq  &  08:26:26  &    &  560  &  F850LP \\ 
   &        &    &  j8fs17pqq  &  08:38:02  &    &  560  &  F850LP \\ 
   &        &    &  j8fs17psq  &  08:50:15  &    &  375  &  F475W  \\ 
   &        &    &  j8fs17puq  &  08:58:46  &    &  375  &  F475W  \\ 
18 &  1692  &  2003-07-13  &  j8fs18tnq  &  07:06:54  &  109.80  &   90  &  F850LP \\ 
   &        &    &  j8fs18toq  &  07:10:42  &    &  560  &  F850LP \\ 
   &        &    &  j8fs18tqq  &  07:22:18  &    &  560  &  F850LP \\ 
   &        &    &  j8fs18tsq  &  07:34:31  &    &  375  &  F475W  \\ 
   &        &    &  j8fs18tuq  &  07:43:02  &    &  375  &  F475W  \\ 
19 &  1030  &  2003-01-31  &  j8fs19ucq  &  20:59:27  &  290.19  &   90  &  F850LP \\ 
   &        &    &  j8fs19udq  &  21:03:15  &    &  560  &  F850LP \\ 
   &        &    &  j8fs19ufq  &  21:14:51  &    &  560  &  F850LP \\ 
   &        &    &  j8fs19uhq  &  21:27:03  &    &  375  &  F475W  \\ 
   &        &    &  j8fs19ujq  &  21:35:35  &    &  375  &  F475W  \\ 
20 &  2000  &  2003-06-21  &  j8fs20j8q  &  06:52:14  &  124.00  &   90  &  F850LP \\ 
   &        &    &  j8fs20j9q  &  06:56:02  &    &  560  &  F850LP \\ 
   &        &    &  j8fs20jbq  &  07:07:38  &    &  560  &  F850LP \\ 
   &        &    &  j8fs20jdq  &  07:19:51  &    &  375  &  F475W  \\ 
   &        &    &  j8fs20jfq  &  07:28:22  &    &  375  &  F475W  \\ 
21 &   685  &  2003-06-13  &  j8fs21mcq  &  17:58:47  &  116.47  &   90  &  F850LP \\ 
   &        &    &  j8fs21mdq  &  18:02:35  &    &  560  &  F850LP \\ 
   &        &    &  j8fs21mfq  &  18:14:11  &    &  560  &  F850LP \\ 
   &        &    &  j8fs21mhq  &  18:26:24  &    &  375  &  F475W  \\ 
   &        &    &  j8fs21mjq  &  18:34:55  &    &  375  &  F475W  \\ 
22 &  1664  &  2003-07-07  &  j8fs22etq  &  11:54:55  &  111.15  &   90  &  F850LP \\ 
   &        &    &  j8fs22euq  &  11:58:43  &    &  560  &  F850LP \\ 
   &        &    &  j8fs22ewq  &  12:10:19  &    &  560  &  F850LP \\ 
   &        &    &  j8fs22eyq  &  12:22:32  &    &  375  &  F475W  \\ 
   &        &    &  j8fs22f0q  &  12:31:03  &    &  375  &  F475W  \\ 
23 &   654  &  2003-06-10  &  j8fs23mnq  &  21:09:46  &  120.01  &   90  &  F850LP \\ 
   &        &    &  j8fs23moq  &  21:13:34  &    &  560  &  F850LP \\ 
   &        &    &  j8fs23mqq  &  21:25:10  &    &  560  &  F850LP \\ 
   &        &    &  j8fs23msq  &  21:37:23  &    &  375  &  F475W  \\ 
   &        &    &  j8fs23muq  &  21:45:54  &    &  375  &  F475W  \\ 
24 &   944  &  2003-06-12  &  j8fs24zpq  &  13:11:28  &  116.00  &   90  &  F850LP \\ 
   &        &    &  j8fs24zqq  &  13:15:16  &    &  560  &  F850LP \\ 
   &        &    &  j8fs24zsq  &  13:26:52  &    &  560  &  F850LP \\ 
   &        &    &  j8fs24zuq  &  13:39:05  &    &  375  &  F475W  \\ 
   &        &    &  j8fs24zwq  &  13:47:36  &    &  375  &  F475W  \\ 
25 &  1938  &  2003-03-05  &  j8fs25cnq  &  21:30:57  &  249.98  &   90  &  F850LP \\ 
   &        &    &  j8fs25coq  &  21:34:45  &    &  560  &  F850LP \\ 
   &        &    &  j8fs25cqq  &  21:46:21  &    &  560  &  F850LP \\ 
   &        &    &  j8fs25csq  &  21:58:33  &    &  375  &  F475W  \\ 
   &        &    &  j8fs25cuq  &  22:07:05  &    &  375  &  F475W  \\ 
26 &  1279  &  2003-07-09  &  j8fs26tiq  &  08:41:47  &  110.32  &   90  &  F850LP \\ 
   &        &    &  j8fs26tjq  &  08:45:35  &    &  560  &  F850LP \\ 
   &        &    &  j8fs26tlq  &  08:57:11  &    &  560  &  F850LP \\ 
   &        &    &  j8fs26tnq  &  09:09:24  &    &  375  &  F475W  \\ 
   &        &    &  j8fs26tpq  &  09:17:55  &    &  375  &  F475W  \\ 
27 &  1720  &  2003-05-15  &  j8fs27a9q  &  08:08:29  &  129.00  &   90  &  F850LP \\ 
   &        &    &  j8fs27aaq  &  08:12:17  &    &  560  &  F850LP \\ 
   &        &    &  j8fs27acq  &  08:23:53  &    &  560  &  F850LP \\ 
   &        &    &  j8fs27aeq  &  08:36:06  &    &  375  &  F475W  \\ 
   &        &    &  j8fs27agq  &  08:44:37  &    &  375  &  F475W  \\ 
28 &   355  &  2003-06-13  &  j8fs28jrq  &  09:58:40  &  116.03  &   90  &  F850LP \\ 
   &        &    &  j8fs28jsq  &  10:02:28  &    &  560  &  F850LP \\ 
   &        &    &  j8fs28juq  &  10:14:04  &    &  560  &  F850LP \\ 
   &        &    &  j8fs28jwq  &  10:26:17  &    &  375  &  F475W  \\ 
   &        &    &  j8fs28jyq  &  10:34:48  &    &  375  &  F475W  \\ 
29 &  1619  &  2003-07-18  &  j8fs29gcq  &  08:45:12  &  110.61  &   90  &  F850LP \\ 
   &        &    &  j8fs29gdq  &  08:49:00  &    &  560  &  F850LP \\ 
   &        &    &  j8fs29gfq  &  09:00:36  &    &  560  &  F850LP \\ 
   &        &    &  j8fs29ghq  &  09:12:49  &    &  375  &  F475W  \\ 
   &        &    &  j8fs29gjq  &  09:21:20  &    &  375  &  F475W  \\ 
30 &  1883  &  2003-03-19  &  j8fs30t7q  &  17:00:03  &  239.94  &   90  &  F850LP \\ 
   &        &    &  j8fs30t8q  &  17:03:51  &    &  560  &  F850LP \\ 
   &        &    &  j8fs30taq  &  17:15:27  &    &  560  &  F850LP \\ 
   &        &    &  j8fs30tcq  &  17:27:40  &    &  375  &  F475W  \\ 
   &        &    &  j8fs30teq  &  17:36:11  &    &  375  &  F475W  \\ 
31 &  1242  &  2003-03-20  &  j8fs31awq  &  17:00:56  &  179.98  &   90  &  F850LP \\ 
   &        &    &  j8fs31axq  &  17:04:44  &    &  560  &  F850LP \\ 
   &        &    &  j8fs31azq  &  17:16:20  &    &  560  &  F850LP \\ 
   &        &    &  j8fs31b1q  &  17:28:33  &    &  375  &  F475W  \\ 
   &        &    &  j8fs31b3q  &  17:37:04  &    &  375  &  F475W  \\ 
32 &   784  &  2003-06-13  &  j8fs32luq  &  16:22:59  &  111.61  &   90  &  F850LP \\ 
   &        &    &  j8fs32lvq  &  16:26:47  &    &  560  &  F850LP \\ 
   &        &    &  j8fs32lxq  &  16:38:23  &    &  560  &  F850LP \\ 
   &        &    &  j8fs32lzq  &  16:50:36  &    &  375  &  F475W  \\ 
   &        &    &  j8fs32m1q  &  16:59:07  &    &  375  &  F475W  \\ 
33 &  1537  &  2003-01-20  &  j8fs33flq  &  19:12:52  &  286.93  &   90  &  F850LP \\ 
   &        &    &  j8fs33fmq  &  19:16:40  &    &  560  &  F850LP \\ 
   &        &    &  j8fs33foq  &  19:28:16  &    &  560  &  F850LP \\ 
   &        &    &  j8fs33fqq  &  19:40:30  &    &  375  &  F475W  \\ 
   &        &    &  j8fs33fsq  &  19:49:00  &    &  375  &  F475W  \\ 
34 &   778  &  2003-06-16  &  j8fs34elq  &  18:00:11  &  130.20  &   90  &  F850LP \\ 
   &        &    &  j8fs34emq  &  18:03:59  &    &  560  &  F850LP \\ 
   &        &    &  j8fs34eoq  &  18:15:35  &    &  560  &  F850LP \\ 
   &        &    &  j8fs34eqq  &  18:27:48  &    &  375  &  F475W  \\ 
   &        &    &  j8fs34esq  &  18:36:19  &    &  375  &  F475W  \\ 
35 &  1321  &  2003-07-16  &  j8fs35seq  &  05:32:39  &  106.93  &   90  &  F850LP \\ 
   &        &    &  j8fs35sfq  &  05:36:27  &    &  560  &  F850LP \\ 
   &        &    &  j8fs35shq  &  05:48:03  &    &  560  &  F850LP \\ 
   &        &    &  j8fs35sjq  &  06:00:16  &    &  375  &  F475W  \\ 
   &        &    &  j8fs35slq  &  06:08:47  &    &  375  &  F475W  \\ 
36 &   828  &  2003-07-17  &  j8fs36y3s  &  05:32:16  &  106.41  &   90  &  F850LP \\ 
   &        &    &  j8fs36y4q  &  05:36:04  &    &  560  &  F850LP \\ 
   &        &    &  j8fs36y6q  &  05:47:40  &    &  560  &  F850LP \\ 
   &        &    &  j8fs36y8q  &  05:59:53  &    &  375  &  F475W  \\ 
   &        &    &  j8fs36yaq  &  06:08:24  &    &  375  &  F475W  \\ 
37 &  1250  &  2003-01-20  &  j8fs37fwq  &  20:48:46  &  286.35  &   90  &  F850LP \\ 
   &        &    &  j8fs37fxq  &  20:52:34  &    &  560  &  F850LP \\ 
   &        &    &  j8fs37fzq  &  21:04:10  &    &  560  &  F850LP \\ 
   &        &    &  j8fs37g1q  &  21:16:22  &    &  375  &  F475W  \\ 
   &        &    &  j8fs37g3q  &  21:24:54  &    &  375  &  F475W  \\ 
38 &  1630  &  2003-07-30  &  j8fs38s0q  &  05:39:01  &  100.01  &   90  &  F850LP \\ 
   &        &    &  j8fs38s1q  &  05:42:49  &    &  560  &  F850LP \\ 
   &        &    &  j8fs38s3q  &  05:54:25  &    &  560  &  F850LP \\ 
   &        &    &  j8fs38s5q  &  06:06:38  &    &  375  &  F475W  \\ 
   &        &    &  j8fs38s7q  &  06:15:09  &    &  375  &  F475W  \\ 
39 &  1146  &  2003-12-28  &  j8fs39cbq  &  22:19:27  &  285.99  &   90  &  F850LP \\ 
   &        &    &  j8fs39ccq  &  22:23:15  &    &  560  &  F850LP \\ 
   &        &    &  j8fs39ceq  &  22:34:51  &    &  560  &  F850LP \\ 
   &        &    &  j8fs39cgq  &  22:47:04  &    &  375  &  F475W  \\ 
   &        &    &  j8fs39ciq  &  22:55:35  &    &  375  &  F475W  \\ 
40 &  1025  &  2003-04-05  &  j8fs40g6q  &  17:13:05  &  174.99  &   90  &  F850LP \\ 
   &        &    &  j8fs40g7q  &  17:16:53  &    &  560  &  F850LP \\ 
   &        &    &  j8fs40g9q  &  17:28:29  &    &  560  &  F850LP \\ 
   &        &    &  j8fs40gbq  &  17:40:42  &    &  375  &  F475W  \\ 
   &        &    &  j8fs40gdq  &  17:49:13  &    &  375  &  F475W  \\ 
41 &  1303  &  2003-07-09  &  j8fs41txq  &  10:17:26  &  114.60  &   90  &  F850LP \\ 
   &        &    &  j8fs41tyq  &  10:21:14  &    &  560  &  F850LP \\ 
   &        &    &  j8fs41u0q  &  10:32:50  &    &  560  &  F850LP \\ 
   &        &    &  j8fs41u2q  &  10:45:03  &    &  375  &  F475W  \\ 
   &        &    &  j8fs41u4q  &  10:53:34  &    &  375  &  F475W  \\ 
42 &  1913  &  2003-04-28  &  j8fs42kuq  &  19:05:58  &  131.00  &   90  &  F850LP \\ 
   &        &    &  j8fs42kvq  &  19:09:46  &    &  560  &  F850LP \\ 
   &        &    &  j8fs42kxq  &  19:21:22  &    &  560  &  F850LP \\ 
   &        &    &  j8fs42kzq  &  19:33:35  &    &  375  &  F475W  \\ 
   &        &    &  j8fs42l1q  &  19:42:06  &    &  375  &  F475W  \\ 
43 &  1327  &  2003-06-12  &  j8fs43aqq  &  16:23:25  &  116.61  &   90  &  F850LP \\ 
   &        &    &  j8fs43arq  &  16:27:13  &    &  560  &  F850LP \\ 
   &        &    &  j8fs43atq  &  16:38:49  &    &  560  &  F850LP \\ 
   &        &    &  j8fs43avq  &  16:51:02  &    &  375  &  F475W  \\ 
   &        &    &  j8fs43axq  &  16:59:33  &    &  375  &  F475W  \\ 
44 &  1125  &  2003-06-11  &  j8fs44qbq  &  09:58:52  &  117.01  &   90  &  F850LP \\ 
   &        &    &  j8fs44qcq  &  10:02:40  &    &  560  &  F850LP \\ 
   &        &    &  j8fs44qeq  &  10:14:16  &    &  560  &  F850LP \\ 
   &        &    &  j8fs44qgq  &  10:26:29  &    &  375  &  F475W  \\ 
   &        &    &  j8fs44qiq  &  10:35:00  &    &  375  &  F475W  \\ 
45 &  1475  &  2003-06-12  &  j8fs45b8q  &  17:59:01  &  117.21  &   90  &  F850LP \\ 
   &        &    &  j8fs45b9q  &  18:02:49  &    &  560  &  F850LP \\ 
   &        &    &  j8fs45bbq  &  18:14:25  &    &  560  &  F850LP \\ 
   &        &    &  j8fs45bdq  &  18:26:38  &    &  375  &  F475W  \\ 
   &        &    &  j8fs45bfq  &  18:35:09  &    &  375  &  F475W  \\ 
46 &  1178  &  2003-01-25  &  j8fs46hjq  &  20:53:34  &  287.17  &   90  &  F850LP \\ 
   &        &    &  j8fs46hkq  &  20:57:22  &    &  560  &  F850LP \\ 
   &        &    &  j8fs46hmq  &  21:08:58  &    &  560  &  F850LP \\ 
   &        &    &  j8fs46hoq  &  21:21:11  &    &  375  &  F475W  \\ 
   &        &    &  j8fs46hqq  &  21:29:42  &    &  375  &  F475W  \\ 
47 &  1283  &  2002-12-25  &  j8fs47tkq  &  18:39:12  &  301.00  &   90  &  F850LP \\ 
   &        &    &  j8fs47tlq  &  18:43:00  &    &  560  &  F850LP \\ 
   &        &    &  j8fs47tnq  &  18:54:36  &    &  560  &  F850LP \\ 
   &        &    &  j8fs47tpq  &  19:06:49  &    &  375  &  F475W  \\ 
   &        &    &  j8fs47trq  &  19:15:20  &    &  375  &  F475W  \\ 
48 &  1261  &  2003-07-13  &  j8fs48u4q  &  08:42:58  &  111.61  &   90  &  F850LP \\ 
   &        &    &  j8fs48u5q  &  08:46:46  &    &  560  &  F850LP \\ 
   &        &    &  j8fs48u7q  &  08:58:22  &    &  560  &  F850LP \\ 
   &        &    &  j8fs48u9q  &  09:10:35  &    &  375  &  F475W  \\ 
   &        &    &  j8fs48ubq  &  09:19:06  &    &  375  &  F475W  \\ 
49 &   698  &  2003-06-21  &  j8fs49jpq  &  08:26:52  &  128.00  &   90  &  F850LP \\ 
   &        &    &  j8fs49jqq  &  08:30:40  &    &  560  &  F850LP \\ 
   &        &    &  j8fs49jsq  &  08:42:16  &    &  560  &  F850LP \\ 
   &        &    &  j8fs49juq  &  08:54:29  &    &  375  &  F475W  \\ 
   &        &    &  j8fs49jwq  &  09:03:00  &    &  375  &  F475W  \\ 
50 &  1422  &  2003-07-09  &  j8fs50ueq  &  11:55:43  &  112.21  &   90  &  F850LP \\ 
   &        &    &  j8fs50ufq  &  11:59:31  &    &  560  &  F850LP \\ 
   &        &    &  j8fs50uhq  &  12:11:07  &    &  560  &  F850LP \\ 
   &        &    &  j8fs50ujq  &  12:23:20  &    &  375  &  F475W  \\ 
   &        &    &  j8fs50ulq  &  12:31:51  &    &  375  &  F475W  \\ 
51 &  2048  &  2003-06-08  &  j8fs51tsq  &  14:46:55  &  131.80  &   90  &  F850LP \\ 
   &        &    &  j8fs51ttq  &  14:50:43  &    &  560  &  F850LP \\ 
   &        &    &  j8fs51tvq  &  15:02:19  &    &  560  &  F850LP \\ 
   &        &    &  j8fs51txq  &  15:14:32  &    &  375  &  F475W  \\ 
   &        &    &  j8fs51tzq  &  15:23:03  &    &  375  &  F475W  \\ 
52 &  1871  &  2003-07-11  &  j8fs52epq  &  07:07:02  &  111.41  &   90  &  F850LP \\ 
   &        &    &  j8fs52eqq  &  07:10:50  &    &  560  &  F850LP \\ 
   &        &    &  j8fs52esq  &  07:22:26  &    &  560  &  F850LP \\ 
   &        &    &  j8fs52euq  &  07:34:39  &    &  375  &  F475W  \\ 
   &        &    &  j8fs52ewq  &  07:43:10  &    &  375  &  F475W  \\ 
53 &     9  &  2003-01-21  &  j8fs53n9q  &  19:12:36  &  265.78  &   90  &  F850LP \\ 
   &        &    &  j8fs53naq  &  19:16:24  &    &  560  &  F850LP \\ 
   &        &    &  j8fs53ncq  &  19:28:00  &    &  560  &  F850LP \\ 
   &        &    &  j8fs53neq  &  19:40:14  &    &  375  &  F475W  \\ 
   &        &    &  j8fs53ngq  &  19:48:44  &    &  375  &  F475W  \\ 
54 &   575  &  2003-07-13  &  j8fs54ulq  &  10:18:14  &  110.61  &   90  &  F850LP \\ 
   &        &    &  j8fs54umq  &  10:22:02  &    &  560  &  F850LP \\ 
   &        &    &  j8fs54uoq  &  10:33:38  &    &  560  &  F850LP \\ 
   &        &    &  j8fs54uqq  &  10:45:51  &    &  375  &  F475W  \\ 
   &        &    &  j8fs54usq  &  10:54:22  &    &  375  &  F475W  \\ 
55 &  1910  &  2003-01-31  &  j8fs55t3q  &  17:48:02  &  283.02  &   90  &  F850LP \\ 
   &        &    &  j8fs55t4q  &  17:51:50  &    &  560  &  F850LP \\ 
   &        &    &  j8fs55t6q  &  18:03:26  &    &  560  &  F850LP \\ 
   &        &    &  j8fs55t8q  &  18:15:39  &    &  375  &  F475W  \\ 
   &        &    &  j8fs55taq  &  18:24:10  &    &  375  &  F475W  \\ 
56 &  1049  &  2003-02-12  &  j8fs56t5q  &  14:47:23  &  289.99  &   90  &  F850LP \\ 
   &        &    &  j8fs56t6q  &  14:51:11  &    &  560  &  F850LP \\ 
   &        &    &  j8fs56t8q  &  15:02:47  &    &  560  &  F850LP \\ 
   &        &    &  j8fs56taq  &  15:15:00  &    &  375  &  F475W  \\ 
   &        &    &  j8fs56tcq  &  15:23:31  &    &  375  &  F475W  \\ 
57 &   856  &  2003-07-11  &  j8fs57f8q  &  08:41:59  &  109.21  &   90  &  F850LP \\ 
   &        &    &  j8fs57f9q  &  08:45:47  &    &  560  &  F850LP \\ 
   &        &    &  j8fs57fbq  &  08:57:23  &    &  560  &  F850LP \\ 
   &        &    &  j8fs57fdq  &  09:09:36  &    &  375  &  F475W  \\ 
   &        &    &  j8fs57ffq  &  09:18:07  &    &  375  &  F475W  \\ 
58 &   140  &  2003-06-15  &  j8fs58beq  &  09:59:11  &  115.61  &   90  &  F850LP \\ 
   &        &    &  j8fs58bfq  &  10:02:59  &    &  560  &  F850LP \\ 
   &        &    &  j8fs58bhq  &  10:14:35  &    &  560  &  F850LP \\ 
   &        &    &  j8fs58bjq  &  10:26:48  &    &  375  &  F475W  \\ 
   &        &    &  j8fs58blq  &  10:35:19  &    &  375  &  F475W  \\ 
59 &  1355  &  2003-01-21  &  j8fs59kcq  &  16:01:59  &  285.20  &   90  &  F850LP \\ 
   &        &    &  j8fs59kdq  &  16:05:47  &    &  560  &  F850LP \\ 
   &        &    &  j8fs59kfq  &  16:17:23  &    &  560  &  F850LP \\ 
   &        &    &  j8fs59khq  &  16:29:36  &    &  375  &  F475W  \\ 
   &        &    &  j8fs59kjq  &  16:38:07  &    &  375  &  F475W  \\ 
60 &  1087  &  2003-01-31  &  j8fs60txq  &  19:23:16  &  280.79  &   90  &  F850LP \\ 
   &        &    &  j8fs60tyq  &  19:27:04  &    &  560  &  F850LP \\ 
   &        &    &  j8fs60u0q  &  19:38:40  &    &  560  &  F850LP \\ 
   &        &    &  j8fs60u2q  &  19:50:53  &    &  375  &  F475W  \\ 
   &        &    &  j8fs60u4q  &  19:59:24  &    &  375  &  F475W  \\ 
61 &  1297  &  2003-02-25  &  j8fs61n8q  &  14:58:09  &  243.38  &   90  &  F850LP \\ 
   &        &    &  j8fs61n9q  &  15:01:57  &    &  560  &  F850LP \\ 
   &        &    &  j8fs61nbq  &  15:13:33  &    &  560  &  F850LP \\ 
   &        &    &  j8fs61ndq  &  15:25:46  &    &  375  &  F475W  \\ 
   &        &    &  j8fs61nfq  &  15:34:17  &    &  375  &  F475W  \\ 
62 &  1861  &  2003-11-20  &  j8fs62lbq  &  23:52:41  &  300.80  &   90  &  F850LP \\ 
   &        &    &  j8fs62lcq  &  23:56:29  &    &  560  &  F850LP \\ 
   &        &  2003-11-21  &  j8fs62leq  &  00:08:05  &    &  560  &  F850LP \\ 
   &        &    &  j8fs62lgq  &  00:20:18  &    &  375  &  F475W  \\ 
   &        &    &  j8fs62liq  &  00:28:49  &    &  375  &  F475W  \\ 
63 &   543  &  2003-02-25  &  j8fs63mrq  &  13:21:25  &  249.98  &   90  &  F850LP \\ 
   &        &    &  j8fs63msq  &  13:25:13  &    &  560  &  F850LP \\ 
   &        &    &  j8fs63muq  &  13:36:49  &    &  560  &  F850LP \\ 
   &        &    &  j8fs63mwq  &  13:49:02  &    &  375  &  F475W  \\ 
   &        &    &  j8fs63myq  &  13:57:33  &    &  375  &  F475W  \\ 
64 &  1431  &  2003-01-10  &  j8fs64ccq  &  23:45:46  &  300.00  &   90  &  F850LP \\ 
   &        &    &  j8fs64cdq  &  23:49:34  &    &  560  &  F850LP \\ 
   &        &  2003-01-11  &  j8fs64cfq  &  00:01:10  &    &  560  &  F850LP \\ 
   &        &    &  j8fs64chq  &  00:13:22  &    &  375  &  F475W  \\ 
   &        &    &  j8fs64cjq  &  00:21:54  &    &  375  &  F475W  \\ 
65 &  1528  &  2003-01-22  &  j8fs65whq  &  19:15:22  &  285.27  &   90  &  F850LP \\ 
   &        &    &  j8fs65wiq  &  19:19:10  &    &  560  &  F850LP \\ 
   &        &    &  j8fs65wkq  &  19:30:46  &    &  560  &  F850LP \\ 
   &        &    &  j8fs65wmq  &  19:42:59  &    &  375  &  F475W  \\ 
   &        &    &  j8fs65woq  &  19:51:30  &    &  375  &  F475W  \\ 
66 &  1695  &  2003-02-01  &  j8fs66clq  &  19:24:47  &  281.96  &   90  &  F850LP \\ 
   &        &    &  j8fs66cmq  &  19:28:35  &    &  560  &  F850LP \\ 
   &        &    &  j8fs66coq  &  19:40:11  &    &  560  &  F850LP \\ 
   &        &    &  j8fs66cqq  &  19:52:24  &    &  375  &  F475W  \\ 
   &        &    &  j8fs66csq  &  20:00:55  &    &  375  &  F475W  \\ 
67 &  1833  &  2003-03-20  &  j8fs67baq  &  18:37:57  &  213.63  &   90  &  F850LP \\ 
   &        &    &  j8fs67bbq  &  18:41:45  &    &  560  &  F850LP \\ 
   &        &    &  j8fs67bdq  &  18:53:21  &    &  560  &  F850LP \\ 
   &        &    &  j8fs67bfq  &  19:05:34  &    &  375  &  F475W  \\ 
   &        &    &  j8fs67bhq  &  19:14:05  &    &  375  &  F475W  \\ 
68 &   437  &  2003-07-15  &  j8fs68mhq  &  10:19:56  &  106.81  &   90  &  F850LP \\ 
   &        &    &  j8fs68miq  &  10:23:44  &    &  560  &  F850LP \\ 
   &        &    &  j8fs68mkq  &  10:35:20  &    &  560  &  F850LP \\ 
   &        &    &  j8fs68mmq  &  10:47:33  &    &  375  &  F475W  \\ 
   &        &    &  j8fs68moq  &  10:56:04  &    &  375  &  F475W  \\ 
69 &  2019  &  2003-03-19  &  j8fs69tnq  &  18:37:08  &  224.18  &   90  &  F850LP \\ 
   &        &    &  j8fs69toq  &  18:40:56  &    &  560  &  F850LP \\ 
   &        &    &  j8fs69tqq  &  18:52:33  &    &  560  &  F850LP \\ 
   &        &    &  j8fs69tsq  &  19:04:46  &    &  375  &  F475W  \\ 
   &        &    &  j8fs69tuq  &  19:13:17  &    &  375  &  F475W  \\ 
70 &    33  &  2003-07-15  &  j8fs70l0q  &  05:30:55  &  107.56  &   90  &  F850LP \\ 
   &        &    &  j8fs70l1q  &  05:34:43  &    &  560  &  F850LP \\ 
   &        &    &  j8fs70l3q  &  05:46:19  &    &  560  &  F850LP \\ 
   &        &    &  j8fs70l5q  &  05:58:32  &    &  375  &  F475W  \\ 
   &        &    &  j8fs70l7q  &  06:07:03  &    &  375  &  F475W  \\ 
71 &   200  &  2003-06-11  &  j8fs71qxq  &  11:34:02  &  115.21  &   90  &  F850LP \\ 
   &        &    &  j8fs71qyq  &  11:37:50  &    &  560  &  F850LP \\ 
   &        &    &  j8fs71r0q  &  11:49:26  &    &  560  &  F850LP \\ 
   &        &    &  j8fs71r2q  &  12:01:39  &    &  375  &  F475W  \\ 
   &        &    &  j8fs71r4q  &  12:10:10  &    &  375  &  F475W  \\ 
72 &   571  &  2003-06-23  &  j8fs72d0q  &  11:40:35  &  119.80  &   90  &  F850LP \\ 
   &        &    &  j8fs72d1q  &  11:44:23  &    &  560  &  F850LP \\ 
   &        &    &  j8fs72d3q  &  11:55:59  &    &  560  &  F850LP \\ 
   &        &    &  j8fs72d5q  &  12:08:12  &    &  375  &  F475W  \\ 
   &        &    &  j8fs72d7q  &  12:16:43  &    &  375  &  F475W  \\ 
73 &    21  &  2003-05-20  &  j8fs73nrq  &  14:34:15  &  131.80  &   90  &  F850LP \\ 
   &        &    &  j8fs73nsq  &  14:38:03  &    &  560  &  F850LP \\ 
   &        &    &  j8fs73nuq  &  14:49:39  &    &  560  &  F850LP \\ 
   &        &    &  j8fs73nwq  &  15:01:52  &    &  375  &  F475W  \\ 
   &        &    &  j8fs73nyq  &  15:10:23  &    &  375  &  F475W  \\ 
74 &  1488  &  2003-07-13  &  j8fs74x8q  &  19:55:08  &  113.01  &   90  &  F850LP \\ 
   &        &    &  j8fs74x9q  &  19:58:56  &    &  560  &  F850LP \\ 
   &        &    &  j8fs74xbq  &  20:10:32  &    &  560  &  F850LP \\ 
   &        &    &  j8fs74xdq  &  20:22:45  &    &  375  &  F475W  \\ 
   &        &    &  j8fs74xfq  &  20:31:16  &    &  375  &  F475W  \\ 
75 &  1779  &  2003-07-19  &  j8fs75lwq  &  08:46:03  &  104.01  &   90  &  F850LP \\ 
   &        &    &  j8fs75lxq  &  08:49:51  &    &  560  &  F850LP \\ 
   &        &    &  j8fs75lzq  &  09:01:27  &    &  560  &  F850LP \\ 
   &        &    &  j8fs75m1q  &  09:13:40  &    &  375  &  F475W  \\ 
   &        &    &  j8fs75m3q  &  09:22:11  &    &  375  &  F475W  \\ 
76 &  1895  &  2003-07-12  &  j8fs76m2q  &  11:57:17  &  108.01  &   90  &  F850LP \\ 
   &        &    &  j8fs76m3q  &  12:01:05  &    &  560  &  F850LP \\ 
   &        &    &  j8fs76m5q  &  12:12:41  &    &  560  &  F850LP \\ 
   &        &    &  j8fs76m7q  &  12:24:54  &    &  375  &  F475W  \\ 
   &        &    &  j8fs76m9q  &  12:33:25  &    &  375  &  F475W  \\ 
77 &  1499  &  2003-07-14  &  j8fs77b4q  &  03:55:42  &  110.21  &   90  &  F850LP \\ 
   &        &    &  j8fs77b5q  &  03:59:30  &    &  560  &  F850LP \\ 
   &        &    &  j8fs77b7q  &  04:11:06  &    &  560  &  F850LP \\ 
   &        &    &  j8fs77b9q  &  04:23:19  &    &  375  &  F475W  \\ 
   &        &    &  j8fs77bbq  &  04:31:50  &    &  375  &  F475W  \\ 
78 &  1545  &  2003-07-14  &  j8fs78ccq  &  05:31:40  &  110.61  &   90  &  F850LP \\ 
   &        &    &  j8fs78cdq  &  05:35:28  &    &  560  &  F850LP \\ 
   &        &    &  j8fs78cfq  &  05:47:04  &    &  560  &  F850LP \\ 
   &        &    &  j8fs78chq  &  05:59:17  &    &  375  &  F475W  \\ 
   &        &    &  j8fs78cjq  &  06:07:48  &    &  375  &  F475W  \\ 
79 &  1192  &  2003-06-06  &  j8fs79fwq  &  14:45:00  &  130.00  &   90  &  F850LP \\ 
   &        &    &  j8fs79fxq  &  14:48:48  &    &  560  &  F850LP \\ 
   &        &    &  j8fs79fzq  &  15:00:24  &    &  560  &  F850LP \\ 
   &        &    &  j8fs79g1q  &  15:12:37  &    &  375  &  F475W  \\ 
   &        &    &  j8fs79g3q  &  15:21:08  &    &  375  &  F475W  \\ 
80 &  1857  &  2003-04-17  &  j8fs80lhq  &  10:58:12  &  148.00  &   90  &  F850LP \\ 
   &        &    &  j8fs80liq  &  11:02:00  &    &  560  &  F850LP \\ 
   &        &    &  j8fs80lkq  &  11:13:36  &    &  560  &  F850LP \\ 
   &        &    &  j8fs80lmq  &  11:25:49  &    &  375  &  F475W  \\ 
   &        &    &  j8fs80loq  &  11:34:20  &    &  375  &  F475W  \\ 
81 &  1075  &  2003-07-14  &  j8fs81cxq  &  07:07:09  &  106.61  &   90  &  F850LP \\ 
   &        &    &  j8fs81cyq  &  07:10:57  &    &  560  &  F850LP \\ 
   &        &    &  j8fs81d0q  &  07:22:33  &    &  560  &  F850LP \\ 
   &        &    &  j8fs81d2q  &  07:34:46  &    &  375  &  F475W  \\ 
   &        &    &  j8fs81d4q  &  07:43:17  &    &  375  &  F475W  \\ 
82 &  1948  &  2003-07-14  &  j8fs82deq  &  08:44:04  &  112.21  &   90  &  F850LP \\ 
   &        &    &  j8fs82dfq  &  08:47:52  &    &  560  &  F850LP \\ 
   &        &    &  j8fs82dhq  &  08:59:28  &    &  560  &  F850LP \\ 
   &        &    &  j8fs82djq  &  09:11:41  &    &  375  &  F475W  \\ 
   &        &    &  j8fs82dlq  &  09:20:12  &    &  375  &  F475W  \\ 
83 &  1627  &  2003-01-22  &  j8fs83wzq  &  20:51:27  &  285.77  &   90  &  F850LP \\ 
   &        &    &  j8fs83x0q  &  20:55:15  &    &  560  &  F850LP \\ 
   &        &    &  j8fs83x2q  &  21:06:51  &    &  560  &  F850LP \\ 
   &        &    &  j8fs83x4q  &  21:19:04  &    &  375  &  F475W  \\ 
   &        &    &  j8fs83x6q  &  21:27:35  &    &  375  &  F475W  \\ 
84 &  1440  &  2003-07-13  &  j8fs84v0q  &  11:57:56  &  108.38  &   90  &  F850LP \\ 
   &        &    &  j8fs84v1q  &  12:01:44  &    &  560  &  F850LP \\ 
   &        &    &  j8fs84v3q  &  12:13:20  &    &  560  &  F850LP \\ 
   &        &    &  j8fs84v5q  &  12:25:33  &    &  375  &  F475W  \\ 
   &        &    &  j8fs84v7q  &  12:34:04  &    &  375  &  F475W  \\ 
85 &   230  &  2003-07-15  &  j8fs85llq  &  07:07:01  &  107.41  &   90  &  F850LP \\ 
   &        &    &  j8fs85lmq  &  07:10:49  &    &  560  &  F850LP \\ 
   &        &    &  j8fs85loq  &  07:22:25  &    &  560  &  F850LP \\ 
   &        &    &  j8fs85lqq  &  07:34:38  &    &  375  &  F475W  \\ 
   &        &    &  j8fs85lsq  &  07:43:09  &    &  375  &  F475W  \\ 
86 &  2050  &  2003-03-23  &  j8fs86v1q  &  17:03:34  &  220.38  &   90  &  F850LP \\ 
   &        &    &  j8fs86v2q  &  17:07:22  &    &  560  &  F850LP \\ 
   &        &    &  j8fs86v4q  &  17:18:58  &    &  560  &  F850LP \\ 
   &        &    &  j8fs86v6q  &  17:31:11  &    &  375  &  F475W  \\ 
   &        &    &  j8fs86v8q  &  17:39:42  &    &  375  &  F475W  \\ 
87 &  1993  &  2003-01-31  &  j8fs87rcq  &  14:36:10  &  282.30  &   90  &  F850LP \\ 
   &        &    &  j8fs87rdq  &  14:39:58  &    &  560  &  F850LP \\ 
   &        &    &  j8fs87rfq  &  14:51:34  &    &  560  &  F850LP \\ 
   &        &    &  j8fs87rhq  &  15:03:47  &    &  375  &  F475W  \\ 
   &        &    &  j8fs87rjq  &  15:12:18  &    &  375  &  F475W  \\ 
88 &   751  &  2003-06-19  &  j8fs88xjq  &  08:24:44  &  120.01  &   90  &  F850LP \\ 
   &        &    &  j8fs88xks  &  08:28:32  &    &  560  &  F850LP \\ 
   &        &    &  j8fs88xmq  &  08:40:08  &    &  560  &  F850LP \\ 
   &        &    &  j8fs88xoq  &  08:52:21  &    &  375  &  F475W  \\ 
   &        &    &  j8fs88xqq  &  09:00:52  &    &  375  &  F475W  \\ 
89 &  1828  &  2003-01-31  &  j8fs89scq  &  16:11:59  &  282.24  &   90  &  F850LP \\ 
   &        &    &  j8fs89sdq  &  16:15:47  &    &  560  &  F850LP \\ 
   &        &    &  j8fs89sfq  &  16:27:23  &    &  560  &  F850LP \\ 
   &        &    &  j8fs89shq  &  16:39:36  &    &  375  &  F475W  \\ 
   &        &    &  j8fs89sjq  &  16:48:07  &    &  375  &  F475W  \\ 
90 &   538  &  2003-07-14  &  j8fs90dvq  &  10:18:27  &  110.20  &   90  &  F850LP \\ 
   &        &    &  j8fs90dwq  &  10:22:15  &    &  560  &  F850LP \\ 
   &        &    &  j8fs90dyq  &  10:33:51  &    &  560  &  F850LP \\ 
   &        &    &  j8fs90e0q  &  10:46:04  &    &  375  &  F475W  \\ 
   &        &    &  j8fs90e2q  &  10:54:35  &    &  375  &  F475W  \\ 
91 &  1407  &  2003-07-15  &  j8fs91jgq  &  02:19:49  &  110.81  &   90  &  F850LP \\ 
   &        &    &  j8fs91jhq  &  02:23:37  &    &  560  &  F850LP \\ 
   &        &    &  j8fs91jjq  &  02:35:13  &    &  560  &  F850LP \\ 
   &        &    &  j8fs91jlq  &  02:47:26  &    &  375  &  F475W  \\ 
   &        &    &  j8fs91jnq  &  02:55:57  &    &  375  &  F475W  \\ 
92 &  1886  &  2003-07-15  &  j8fs92klq  &  03:56:27  &  110.81  &   90  &  F850LP \\ 
   &        &    &  j8fs92kmq  &  04:00:15  &    &  560  &  F850LP \\ 
   &        &    &  j8fs92koq  &  04:11:51  &    &  560  &  F850LP \\ 
   &        &    &  j8fs92kqq  &  04:24:04  &    &  375  &  F475W  \\ 
   &        &    &  j8fs92ksq  &  04:32:35  &    &  375  &  F475W  \\ 
93 &  1199  &  2003-02-24  &  j8fs93i3q  &  22:58:17  &  253.79  &   90  &  F850LP \\ 
   &        &    &  j8fs93i4q  &  23:02:05  &    &  560  &  F850LP \\ 
   &        &    &  j8fs93i6q  &  23:13:41  &    &  560  &  F850LP \\ 
   &        &    &  j8fs93i8q  &  23:25:54  &    &  375  &  F475W  \\ 
   &        &    &  j8fs93iaq  &  23:34:25  &    &  375  &  F475W  \\ 
94 &  1743  &  2003-07-18  &  j8fs94grq  &  10:21:12  &  107.41  &   90  &  F850LP \\ 
   &        &    &  j8fs94gsq  &  10:25:00  &    &  560  &  F850LP \\ 
   &        &    &  j8fs94guq  &  10:36:36  &    &  560  &  F850LP \\ 
   &        &    &  j8fs94gwq  &  10:48:49  &    &  375  &  F475W  \\ 
   &        &    &  j8fs94gyq  &  10:57:20  &    &  375  &  F475W  \\ 
95 &  1539  &  2003-06-24  &  j8fs95jgq  &  08:29:37  &  113.81  &   90  &  F850LP \\ 
   &        &    &  j8fs95jhq  &  08:33:25  &    &  560  &  F850LP \\ 
   &        &    &  j8fs95jjq  &  08:45:01  &    &  560  &  F850LP \\ 
   &        &    &  j8fs95jlq  &  08:57:14  &    &  375  &  F475W  \\ 
   &        &    &  j8fs95jnq  &  09:05:45  &    &  375  &  F475W  \\ 
96 &  1185  &  2003-07-11  &  j8fs96fnq  &  10:18:30  &  114.21  &   90  &  F850LP \\ 
   &        &    &  j8fs96foq  &  10:22:18  &    &  560  &  F850LP \\ 
   &        &    &  j8fs96fqq  &  10:33:54  &    &  560  &  F850LP \\ 
   &        &    &  j8fs96fsq  &  10:46:07  &    &  375  &  F475W  \\ 
   &        &    &  j8fs96fuq  &  10:54:38  &    &  375  &  F475W  \\ 
97 &  1826  &  2003-01-21  &  j8fs97lrq  &  17:38:15  &  278.39  &   90  &  F850LP \\ 
   &        &    &  j8fs97lsq  &  17:42:03  &    &  560  &  F850LP \\ 
   &        &    &  j8fs97luq  &  17:53:39  &    &  560  &  F850LP \\ 
   &        &    &  j8fs97lwq  &  18:05:52  &    &  375  &  F475W  \\ 
   &        &    &  j8fs97lyq  &  18:14:23  &    &  375  &  F475W  \\ 
98 &  1512  &  2003-07-19  &  j8fs98kxq  &  05:33:22  &  106.21  &   90  &  F850LP \\ 
   &        &    &  j8fs98kyq  &  05:37:10  &    &  560  &  F850LP \\ 
   &        &    &  j8fs98l0q  &  05:48:46  &    &  560  &  F850LP \\ 
   &        &    &  j8fs98l2q  &  06:00:59  &    &  375  &  F475W  \\ 
   &        &    &  j8fs98l4q  &  06:09:30  &    &  375  &  F475W  \\ 
99 &  1489  &  2003-07-15  &  j8fs99m2q  &  08:43:52  &  111.61  &   90  &  F850LP \\ 
   &        &    &  j8fs99m3q  &  08:47:40  &    &  560  &  F850LP \\ 
   &        &    &  j8fs99m5q  &  08:59:16  &    &  560  &  F850LP \\ 
   &        &    &  j8fs99m7q  &  09:11:29  &    &  375  &  F475W  \\ 
   &        &    &  j8fs99m9q  &  09:20:00  &    &  375  &  F475W  \\ 
100&  1661  &  2003-12-25  &  j8fsa0f4q  &  23:56:43  &  289.99  &   90  &  F850LP \\ 
   &        &  2003-12-26  &  j8fsa0f5q  &  00:00:31  &    &  560  &  F850LP \\ 
   &        &    &  j8fsa0f7q  &  00:12:07  &    &  560  &  F850LP \\ 
   &        &    &  j8fsa0f9q  &  00:24:20  &    &  375  &  F475W  \\ 
   &        &    &  j8fsa0fbq  &  00:32:51  &    &  375  &  F475W  \\ 
\enddata

\end{deluxetable}

\clearpage

\clearpage

\clearpage

\clearpage

\clearpage

\clearpage

\clearpage

\clearpage

\clearpage

\end{document}